\documentclass[11pt]{article}

\usepackage{amsfonts,amssymb,jheppub,dsfont,graphicx,mathrsfs,tikz,physics,bbm}
\usetikzlibrary{decorations.markings}
\usetikzlibrary{decorations.pathreplacing}
\allowdisplaybreaks

\usepackage[numbers,sort&compress]{natbib}

\DeclareMathAlphabet{\dutchcal}{U}{dutchcal}{m}{n}
\SetMathAlphabet{\dutchcal}{bold}{U}{dutchcal}{b}{n}
\DeclareMathAlphabet{\dutchbcal} {U}{dutchcal}{b}{n}

\newcommand{\iu}{\mathrm{i}\mkern1mu}
\newcommand{\chan}{\mathfrak{C}}
\newcommand{\h}{\mathbbm{h}}

\definecolor{back-clr}{RGB}{209, 224, 224}

\title{One- and two-dimensional higher-point conformal blocks as free-particle wavefunctions in  AdS$_3^{\otimes m}$}

\author{Jean-Fran\c{c}ois Fortin$^{\ast}$, Wen-Jie Ma$^{\S,\ddag}$, Sarthak Parikh$^{+}$, Lorenzo Quintavalle$^{\ast}$, and Witold Skiba$^{\dagger}$}

\affiliation{
$^\ast$D\'epartement de Physique, de G\'enie Physique et d'Optique, Universit\'e Laval, Qu\'ebec, QC~G1V~0A6, Canada\\
$^{\S}$Beijing Institute of Mathematical Sciences and Applications (BIMSA), Beijing, 101408, China}
\affiliation{$^{\ddag}$Yau Mathematical Sciences Center (YMSC), Tsinghua University, Beijing, 100084, China}
\affiliation{$^+$Department of Physics, Indian Institute of Technology Delhi, Hauz Khas, New Delhi 110016, India}
\affiliation{$^\dagger$Department of Physics, Yale University, New Haven, CT 06520, USA
}

\emailAdd{jean-francois.fortin@phy.ulaval.ca}
\emailAdd{wenjia.ma@bimsa.cn}
\emailAdd{sarthak@physics.iitd.ac.in}
\emailAdd{lorenzo.quintavalle.1@ulaval.ca}
\emailAdd{witold.skiba@yale.edu}

\abstract{We establish that all of the one- and two-dimensional global conformal blocks are, up to some choice of prefactor, free-particle wavefunctions in tensor products of AdS$_3$ or limits thereof. Our first core observation is that the six-point comb-channel conformal blocks correspond to free-particle wavefunctions on an AdS$_3$ constructed directly in cross-ratio space. This construction generalizes to blocks for a special class of diagrams, which are determined as free-particle wavefunctions in tensor products of AdS$_3$. Conformal blocks for all the remaining topologies are obtained as limits of the free wavefunctions mentioned above. Our results show directly that the integrable models associated with all one- and two-dimensional conformal blocks can be seen as limits of free theory, and manifest a relation between AdS and CFT kinematics that lies outside of the standard AdS/CFT dictionary. We complete the discussion by providing explicit Feynman-like rules that can be used to work out blocks for all topologies, as well as a Mathematica notebook that allows simple computation of Casimir equations and series expansions for blocks, by requiring just an OPE diagram as input.}

\date{October 2023} 

\begin{document}

\maketitle



\section{Introduction}\label{SecIntro}

The framework of quantum field theory (QFT) encompasses successful theories in high-energy physics (most notably the Standard Model of particle physics) and condensed matter physics.  It is however notoriously difficult to understand---let alone solve---strongly-coupled QFTs without another handle, as for example the presence of an extra symmetry.  Such a symmetry exists naturally at the renormalization group flow fixed points of QFTs, where the usual Poincar\'e symmetry group gets enhanced to the conformal symmetry group, leading to conformal field theories (CFTs).  Well-known CFTs in high-energy physics comprise the Banks-Zaks fixed points of quantum chromodynamics as well as their supersymmetric cousins, while in the realm of condensed matter, CFTs correspond to systems that undergo second-order phase transitions such as the ferromagnets described by the Ising model.  CFTs are also a very useful tool in the understanding of quantum gravity through the AdS/CFT correspondence.

The extension from Poincar\'e to the whole conformal symmetry group separates the spectrum of operators into quasi-primaries and descendants and guarantees the existence of a convergent operator product expansion (OPE) \cite{Ferrara:1971vh,Ferrara:1972cq,Ferrara:1973eg,Mack:1976pa}. The OPE re-expresses the product of two operators inserted at two distinct points as an infinite sum over ``exchanged'' quasi-primary operators and their descendants.  Conformal invariance, along with the aforementioned tools, thus allows for an understanding of correlation functions in terms of kinematic quantities---the so-called conformal blocks---and dynamical OPE coefficients.  In this setup, the conformal blocks are completely fixed by conformal invariance while the OPE coefficients act like coupling constants distinguishing between different CFTs.  Demanding associativity of the OPE on correlation functions generates non-trivial constraints on the dynamical inputs, even at strong coupling, eventually fixing (at least partially) the OPE coefficients.  The usual starting point for such an endeavor is with four-point correlation functions, where the associativity constraints correspond to the celebrated bootstrap equations \cite{Ferrara:1973yt,Polyakov:1974gs}.

Four-point conformal blocks of four external scalar quasi-primaries have been studied extensively with different tools by Dolan and Osborn \cite{Dolan:2000ut,Dolan:2003hv}, culminating in their use of the Casimirs as differential eigenoperators of which conformal blocks are eigenfunctions with appropriate asymptotic behavior \cite{Dolan:2011dv}.  Their work thus led to a deeper understanding of scalar conformal blocks, which is invaluable when implementing the bootstrap program.  The latter saw an explosion of interest after the seminal work of \cite{Rattazzi:2008pe} which concretely showed that the conformal bootstrap equations can numerically constrain the dynamical CFT data.  The interested reader is advised to look at the recent reviews~\cite{Poland:2018epd,Hartman:2022zik,Poland:2022qrs} as well as references therein.

To fully leverage the constraining power of the bootstrap or the applicability of CFTs, it is of great importance to consider setups that involve data for operators with spin. A common approach in this sense is to include four-point correlators of spinning operators as part of the bootstrap analysis. 
Although completely fixed by conformal invariance, four-point conformal blocks are nevertheless quite difficult to compute explicitly when the external quasi-primary operators are in non-trivial representations of the Lorentz group, especially if their spin parameters are large. 
For this among other reasons, the idea that certain spinning data may be more easily accessible from the study of higher-point (\textit{a.k.a.}\ multipoint) functions that involve only external scalars has recently started to get more traction, \textit{e.g.}\ in the context of lightcone bootstrap~\cite{Bercini:2020msp,Antunes:2021kmm,Kaviraj:2022wbw}, numerical bootstrap~\cite{Poland:2023vpn}, Regge theory~\cite{Costa:2023wfz}, or the study of constraints from the OPE limit~\cite{Anous:2021caj}. However, to date, results on conformal blocks involving more than four scalars are still limited. In order to better pursue this program, it is then crucial to develop a better understanding of higher-point conformal blocks.

Higher-point conformal blocks in several channels and their associated bootstrap equations have recently been studied in different spacetime dimensions from a variety of techniques. In general $d$ spacetime dimensions, results for scalar-exchange conformal blocks in various topologies have been obtained in~\cite{Rosenhaus:2018zqn,Bhatta:2018gjb,Parikh:2019ygo,Jepsen:2019svc,Parikh:2019dvm,Fortin:2019zkm,Fortin:2020yjz,Anous:2020vtw,Fortin:2020bfq,Hoback:2020pgj,Fortin:2022grf}, including a set of Feynman-like rules to write explicit series expansions of these blocks~\cite{Hoback:2020pgj,Fortin:2022grf}.
For more general exchanged operators, a series representation for five-point conformal blocks has been computed in~\cite{Goncalves:2019znr}, five-point recursion relations have been worked out in~\cite{Poland:2021xjs}, and special limits for five- and six-point blocks have been determined in~\cite{Bercini:2020msp,Antunes:2021kmm,Kaviraj:2022wbw,Costa:2023wfz}. A generalization of the Casimir approach of Dolan and Osborn for arbitrary higher-point blocks has also been introduced in~\cite{Buric:2020dyz,Buric:2021ywo,Buric:2021ttm,Buric:2021kgy}.

Given the complexity of the problem, a good starting point to understand general higher-point conformal blocks is to consider simpler one-dimensional systems that are invariant under the conformal group. These are not only the perfect laboratory to sharpen general CFT tools, \textit{e.g.}\ the Polyakov/functional bootstrap~\cite{Mazac:2018mdx,Mazac:2018ycv,Ghosh:2023lwe} or the Boostrability approach~\cite{Cavaglia:2021bnz,Cavaglia:2022qpg,Cavaglia:2022yvv}, but are also systems directly relevant to higher-dimensional theories when considering correlators restricted to line defects~\cite{Giombi:2017cqn,Liendo:2018ukf,Ferrero:2021bsb,Barrat:2021tpn,Barrat:2022eim}. Given the simpler structure of 1d CFTs, its conformal blocks are much more under control and have been determined using the large central charge limit in CFT$_2$~\cite{Alkalaev:2015fbw}, shadow integrals~\cite{Rosenhaus:2018zqn} and Wilson networks in first-order formalism~\cite{Bhatta:2016hpz,Besken:2016ooo,Alkalaev:2023axo} for the comb-channel case, and via direct application of the OPE in~\cite{Fortin:2020zxw} for arbitrary topologies via a set of Feynman-like rules.

In this paper, we aim to study the one-dimensional higher-point conformal blocks from the perspective of the differential equations they satisfy, which allows a more direct understanding of the space of functions to which these belong. This is in line with the pioneering results of~\cite{Isachenkov:2016gim,Isachenkov:2017qgn}, which determined how the four-point conformal blocks in any dimensions can be interpreted as quantum-mechanical eigenfunctions for the integrable Calogero-Sutherland Hamiltonian. Here, by analyzing the Casimir equations for general higher-point blocks, we will be able to show that all of the one-dimensional higher-point conformal blocks directly correspond to (limits of) free-particle wavefunctions in products of AdS$_3$ spaces. Let us stress that all of the results of this paper also trivially apply to two-dimensional global\footnote{As opposed to two-dimensional local CFTs where the full power of the Virasoro algebra is utilized.  The existence of the Virasoro algebra leads to an extra separation of quasi-primary fields into primary and quasi-primary fields.  Since we will only be concerned by global CFTs, from now on we will not differentiate between them and use primary and quasi-primary interchangeably.} CFTs, as the symmetry group factors in two copies of $SO(2,1)$ and thus the corresponding conformal blocks are just products of two one-dimensional blocks.

The paper is organized as follows: Section~\ref{sec:notation} introduces the notation and some conventions on correlation functions, conformal blocks, and Casimirs that are used throughout the paper.  Section~\ref{SecCasComb} discusses low-point conformal blocks in the comb channel.  We first introduce the Casimir operators for four-, five- and six-point comb topologies, and differentiate between the standard conformal blocks and rescaled cousins which have simple interpretations in the context of integrability, where Casimirs are interpreted as quantum-mechanical Hamiltonians and associated conserved charges.  In particular, we show that one of the three non-trivial and independent Casimirs of six-point comb-channel blocks is nothing else than the Laplacian on AdS$_3$, demonstrating that the (properly normalized) blocks can be identified with free-particle wavefunctions in AdS$_3$ equipped with two extra conserved charges.  Five- and four-point scalar blocks are then understood from the six-point solutions as different limits (OPE limits or identity limits), showing for example that five-point conformal blocks can be seen as charged particles propagating in a constant background magnetic field on AdS$_2$ space.  This new point of view is then generalized to higher-point conformal blocks for arbitrary topologies in Section~\ref{SecGenTop}.  There, we first state our conventions for higher-point correlation functions in arbitrary topologies.  Then we carefully study ``six-point-constructible'' higher-point blocks, \textit{i.e.}\ blocks in specific OPE channels that have the important property of being built directly from six-point comb channels glued together in well-defined ways, leading to the determination of their blocks as freely propagating wavefunctions on products of AdS$_3$.  Interestingly, the gluing procedure can be understood directly at the level of Casimir operators by enforcing identities between some of them.  We also discuss how general higher-point correlation functions can be generated from six-point-constructible diagrams by making use of the OPE and identity limits mentioned above.  We conclude this section with simple Feynman-like rules appropriate for our conventions, that lead straightforwardly to conformal blocks in arbitrary topologies. For a more accessible use of our results, we supplement this work with a \texttt{Mathematica} notebook equipped with efficient functions to compute general one-dimensional Casimir equations and conformal block expressions by simply providing an OPE diagram as input.   
Finally, we conclude and discuss future directions in Section~\ref{SecConc}, followed by an appendix where we show explicitly one non-trivial example in which the limits (OPE and identity) are applied to reach a general topology from a specific six-point-constructible diagram.


\section{Notation and conventions}\label{sec:notation}

\begin{figure}[htp]
    \centering
    \includegraphics{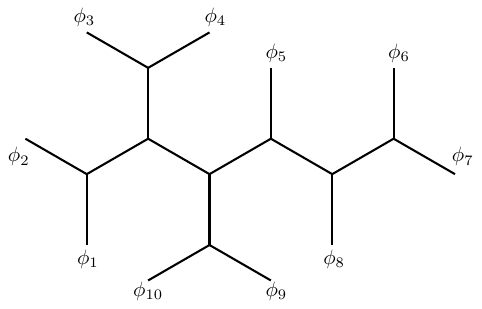}
    \caption{Example of an OPE diagram for ten-point correlators. Every labeled leg represents a field involved in the correlator, and every internal leg stands for the exchange of a tower of conformal multiplets in an OPE.}
    \label{fig:higher_pt}
\end{figure}

The setting of this work is that of correlators of $N$ scalar primary fields $\phi_i(x_i)$ in a one-dimensional global CFT. The results we will derive are trivially extended to two-dimensional global CFTs, as the symmetry group for that case factorizes in two copies of the one-dimensional group. The $\mathrm{SO}(2,1)$ Ward identities combined with the use of a convergent OPE imply that these correlators can be decomposed as follows
\begin{multline}
    \expval{\phi_{1} \phi_{2}\dots \phi_{N}}:= \expval{\phi_{h_1}(x_1) \phi_{h_2}(x_2)\dots \phi_{h_N}(x_N)}=\Omega_N^{\{h_i\}}(\abs{x_{ij}})f(z_1,\dots,z_{N-3})\\ 
    =\Omega_N^{\{h_i\}}(\abs{x_{ij}}) \sum_{\mathbbm{h}_1,\dots,\mathbbm{h}_{N-3}} \left( \prod_{\nu=1}^{N-2} C_{\nu_{(1)}\nu_{(2)}\nu_{(3)}} \right)\tilde{\psi}_{\{\mathbbm{h}_r\}}^{(\chan)}(z_1,\dots,z_{N-3})\,.
    \label{eq:N-point_decomposition}
\end{multline}
In the expression~\eqref{eq:N-point_decomposition}, the prefactor $\Omega_N^{\{h_i\}}$ encodes the covariance properties of the correlator and depends on the distances $x_{ij}=x_i-x_j$ and on the one-dimensional conformal weights of the external fields $h_i$. Once this prefactor is factored out, what remains is an invariant object $f(z_1,\dots,z_{N-3})$ that depends only on $N-3$ conformally-invariant degrees of freedom, the \emph{cross-ratios}~$z_r$. This object can be greatly simplified by performing $N-2$ OPEs, which iteratively replace a product of two operators with an infinite sum of ``exchanged'' operators with different conformal dimensions~$\mathbbm{h}_r$, therefore reducing the correlators to linear combinations of two-point functions. 
The various inequivalent ways in which this procedure can be followed, known as \emph{OPE channels}, are parametrized by binary trees with $N$ leaves such as that of Figure~\ref{fig:higher_pt}.
In this picture, every external line represents an insertion of a scalar field $\phi_i$ in the correlator, and every internal line represents a field $\mathcal{O}_{\mathbbm{h}_r}$ exchanged in an OPE, with conformal dimension $\mathbbm{h}_r$. To establish some conventions, we will use from now on the symbol $\chan$ as a generic label for a choice of OPE channel, and within a channel $\chan$, we will label the external legs with the index $i$, the internal legs with the index $r$, and the three-point vertices with the index $\nu$. When we will have to refer to the three legs attached to the vertex $\nu$ individually, we will use the symbol $\nu_{(j)}$, with $j=1,2,3$. When discussing comb channels properly ordered from field $\phi_1$ first to field $\phi_N$ last, we will often use just a number $(N)$ in place of a label $\chan$ to indicate the OPE channel. Also, when it may be obvious from the context which channel we are talking about, we may avoid writing explicitly the dependence on the channel $\chan$ of the quantities we will analyze.

The procedure we described confines the kinematical dependence of the correlator to some functions $\tilde{\psi}_{\{\mathbbm{h}_r\}}^{(\chan)}$, known as conformal blocks, which act in a loose sense as a basis of functions labeled by the exchanged conformal dimensions, and in terms of which the correlator can be expanded. In one dimension, these functions can be characterized as joint eigenfunctions of $N-3$ quadratic \emph{Casimir operators}, one for every field exchanged in an internal leg. Every Casimir operator is constructed by summing generators in order to act on the tensor-product representation associated with the exchanged field, and contracting two copies of this sum with the symmetric invariant tensor $\kappa_{ab}$ as follows:
\begin{equation}
    \mathcal{C}_r^2=\kappa_{ab} \left(\sum_{i\prec r}\mathcal{T}^a_{(i)}\right)\left(\sum_{i\prec r}\mathcal{T}^b_{(i)}\right).
\end{equation}
In this expression, we indicate with the symbol $i\prec r$ all external fields $\phi_i$ that belong to one of the two parts of the diagram separated by the internal leg $r$, and with $\mathcal{T}^a_{(i)}$ the conformal generators with adjoint index $a$ that act on the field $\phi_i$. Note that the quadratic Casimir operator does not depend on which side of $r$ we consider for the sum, as if one picks the complementary choice of legs $i\succ r$, each factor of generators acquires a minus sign due to the conformal Ward identities $\sum_{i\succ r}\mathcal{T}^a_{(i)}=-\sum_{i\prec r}\mathcal{T}^a_{(i)}$, but the quadratic Casimir remains unaltered.

For every quadratic Casimir operator constructed as above, one can easily obtain a conformally-invariant differential equation for the conformal blocks $\tilde{\psi}_{\{\mathbbm{h}_r\}}^{(\chan)}$ via the eigenvalue equation for $\mathcal{C}_r^2$:

\begin{equation}
    \widetilde{\mathcal{D}}_{r}^{(\chan)}\tilde{\psi}_{\{\mathbbm{h}_r\}}^{(\chan)}:= \frac{1}{\Omega_N^{\{h_i\}}}\mathcal{C}_{r}^2 \left(\Omega_N^{\{h_i\}} \tilde{\psi}_{\{\mathbbm{h}_r\}}^{(\chan)}\right)= \mathbbm{h}_r\left(\mathbbm{h}_r-1\right)\tilde{\psi}_{\{\mathbbm{h}_r\}}^{(\chan)}\,,
    \label{eq:general_Casimir_equation}
\end{equation}
which can be expressed in terms of the cross-ratios $z_r$.\footnote{It is important to note that the quadratic Casimir operators associated with external operators are simply constants when acting on the blocks.} Since the number of (commuting) Casimir operators matches the number of degrees of freedom of the system---the cross-ratios---we know that the problem is integrable, and the conformal blocks can thus be fully determined by solving the Casimir differential equations and by requiring appropriate boundary conditions compatible with the OPE.

\begin{figure}[htp]
    \centering
    \includegraphics{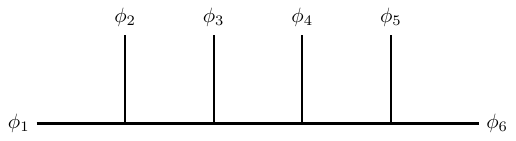}
    \caption{Example of comb-channel OPE diagram for six-point correlators.}
    \label{fig:Comb-six}
\end{figure}

Every choice of OPE channel $\chan$ leads to a different set of Casimir operators $\mathcal{C}_r^2$, but among the possible channels there is one class that stands out for the simplicity of its operators: the \emph{comb channel} expansions, in which every OPE involves at least one external field, see Figure~\ref{fig:Comb-six} for a six-point example.
When considering comb channels, we will label fields in ascending order from left to right within the comb, such that all the Casimir operators acquire the form
\begin{equation}
    \mathcal{C}_{1\dots m}^2=\kappa_{ab} \mathcal{T}_{12\dots m}^a\mathcal{T}_{12\dots m}^b= \kappa_{ab} \mathcal{T}_{(m+1)\dots N}^a\mathcal{T}_{(m+1)\dots N}^b=\mathcal{C}_{m+1\dots N}^2,
\end{equation}
where we used the shorthand
\begin{equation}
    \mathcal{T}_{m\dots n}^a=\left(\sum_{m\le i \le n}\mathcal{T}^a_{(i)}\right).
\end{equation}
We are now ready to study the conformal blocks to uncover their link with (limits of) free-particle wavefunctions in product of AdS$_3$ spaces.


\section{Six-point comb channel and the \texorpdfstring{AdS$_3$}{AdS3} free Hamiltonian}\label{SecCasComb}

In this section, we aim to discuss the simplest cases of comb-channel higher-point conformal blocks, and provide an interpretation of the space these conformal blocks live in. We will start by constructing the $N=4,5,6$ Casimir equations, to then show that one of the six-point Casimir operators is nothing but the free Hamiltonian of AdS$_3$. This means that the corresponding six-point conformal blocks are free-particle wavefunctions in AdS$_3$, and will allow us to construct their expressions in terms of a canonical basis of wave modes that diagonalize two Cartan generators. By taking OPE limits in Section~\ref{ssec:OPE-limits-5-and-4} and/or ``identity limits'' in Section~\ref{ssec:identity_reduction} to reduce the number of external legs, we will show how this interpretation encapsulates also the $N<6$ cases.

\subsection{The Casimir operators for \texorpdfstring{$N\le 6$}{N<=6} comb-channel blocks}\label{ssec:Casimir_expr}
We now proceed to specify some conventions for the conformal block expansions~\eqref{eq:N-point_decomposition}, and to derive the expressions of the Casimir differential operators~\eqref{eq:general_Casimir_equation} that determine the $N$-point blocks we aim to analyze, momentarily focusing only on the $(N\leq6)$-point comb channel. This will be enough to establish, in Section~\ref{SecHamComb}, our starting point for the relation between one-dimensional higher-point blocks and free-particle wavefunctions in AdS$_3$.

When writing a conformal block expansion, while in principle any choice of prefactor $\Omega_N^{\{h_i\}}$ and conformally-invariant cross-ratios $\{z_r\}$ is valid, different choices can affect the complexity of computations and of the final expressions one gets for conformal blocks. Here, we will follow the conventions of~\cite{Rosenhaus:2018zqn}, where one-dimensional higher-point conformal blocks in the comb channel were first determined. 
An extension of these conventions has been shown in~\cite{Buric:2021kgy} to be particularly useful also for higher-dimensional comb-channel blocks, as the Casimir differential operators have coefficients that are polynomial in the cross-ratios.\footnote{This is a feature that seems to be valid whenever the choice of prefactor and cross-ratios respects an ``OPE flow'' as defined in~\cite{Fortin:2022grf} (see also Section~\ref{SecGenTop}) in the context of Feynman rules for scalar conformal blocks.}

Explicitly, this means we are considering correlators of fields $\phi_{h_i}(x_i)$ located at positions $x_1> x_2> \dots >x_N$, and that we express these in terms of the following prefactors and cross-ratios:
\begin{equation}
    \Omega_N^{\{h_i\}}=\left(\frac{x_{23}}{x_{12}x_{13}}\right)^{h_1}\left(\frac{x_{N-2,N-1}}{x_{N-2,N}\,x_{N-1,N}}\right)^{h_N}\prod_{i=1}^{N-2} \left(\frac{x_{i,i+2}}{x_{i,i+1}x_{i+1,i+2}}\right)^{h_{i+1}}, \quad z_r=\frac{x_{r,r+1}x_{r+2,r+3}}{x_{r,r+2}x_{r+1,r+3}}\,,
    \label{eq:prefactor_cr_conventions}
\end{equation}
where $r=1, \ldots, N-3$.
Furthermore, we will use the notation for the following difference of conformal dimensions appearing at the extremes of the comb:
\begin{equation}
    a=h_2-h_1\,, \qquad \tilde{a}=h_{N-1}-h_{N}\,.
\end{equation}
With these notations in hand, we can now proceed to work out the comb-channel Casimir equations~\eqref{eq:general_Casimir_equation} for $N\le 6$, which are a fundamental starting point for the general-channel expressions.


\subsubsection{Four-point Casimir operators}

Since two- and three-point functions of scalar fields are completely determined by conformal symmetry up to an overall constant factor, the first non-trivial case in which conformal blocks appear in the OPE expansion of correlators is that of four-point correlators. This case was studied already starting from the  '70s~\cite{Ferrara:1973vz,Ferrara:1974ny}, but got renewed attention in the early 2000s with the works of Dolan and Osborn~\cite{Dolan:2000ut,Dolan:2003hv,Dolan:2011dv,Osborn:2012vt}, who introduced a technique involving Casimir operators to compute conformal blocks.  In this setting---four points in one dimension---there is only one Casimir operator $\mathcal{C}_{12}^2=\mathcal{C}_{34}^2$ which leads to a differential operator that is directly related to that for Gauss' hypergeometric function. In our conventions, this acquires the form:
\begin{equation}
    \widetilde{\mathcal{D}}_{12}\tilde{\psi}_{\mathbbm{h}}=\left[\left(1-z_1\right)z_1^2\partial_{z_1}^2- \left(1-a-\tilde{a}\right)z_1^2\partial_{z_1}- a \tilde{a} z_1   \right]\tilde{\psi}_{\mathbbm{h}}\,.
\label{eq:Four-pt-GaussHyp-Diff-Eq}
\end{equation}
It was observed in~\cite{Isachenkov:2016gim,Isachenkov:2017qgn} how this operator is intimately related to the integrable P\"oschl-Teller Hamiltonian~\cite{Poschl:1933zz}, as well as how the $d>2$ case is related to the Calogero-Sutherland Hamiltonian for BC$_2$ root system. It is in line with this observation that we will approach our study of the higher-point equations. In our general analysis, we will see how this operator represents an exception, as it will not appear again in any other $N$-point differential equation.


\subsubsection{Five-point Casimir operators}
\label{ssec:5-pt-Casimirs}
\begin{figure}
    \centering
    \includegraphics{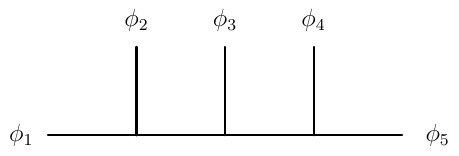}
    \caption{Five-point OPE diagram. The associated conformal blocks can be labeled by eigenvalues of Casimir operators~\eqref{eq:five_pt_Casimirs} associated with the two internal legs of this figure.}
    \label{fig:Five-points_comb}
\end{figure}

In the five-point case, there are two independent Casimirs that can be constructed. Considering the channel in Figure~\ref{fig:Five-points_comb}, we have two relevant Casimir operators:
\begin{equation}
    \mathcal{C}_{12}^2=\kappa_{ab}\mathcal{T}_{12}^a \mathcal{T}_{12}^b\,, \qquad \mathcal{C}_{45}^2=\kappa_{ab}\mathcal{T}_{45}^a \mathcal{T}_{45}^b\,.
    \label{eq:five_pt_Casimirs}
\end{equation}
Using~\eqref{eq:general_Casimir_equation} with our conventions~\eqref{eq:prefactor_cr_conventions}, one obtains
\begin{gather}
    \widetilde{\mathcal{D}}_{12}=z_1\left[(1-z_1)z_1\partial_{z_1}^2-z_1 z_2 \partial_{z_1}\partial_{z_2}-(1-a-h_3) z_1 \partial_{z_1} + a z_2 \partial_{z_2} -a h_3\right],\label{eq:fiveptCas12}\\
    \widetilde{\mathcal{D}}_{45}=z_2\left[(1-z_2)z_2\partial_{z_2}^2-z_1 z_2 \partial_{z_1}\partial_{z_2}-(1-\tilde{a}-h_3) z_2 \partial_{z_2} + \tilde{a} z_1 \partial_{z_1} -\tilde{a} h_3\right].\label{eq:fiveptCas45}
\end{gather}
Note how the $\mathbb{Z}_2$ symmetry of the diagram is manifest in these expressions, corresponding to the action of an operator $\mathcal{P}_5$ that exchanges $z_1\leftrightarrow z_2$ and $a\leftrightarrow \tilde{a}$. 


\subsubsection{Six-point (comb) Casimir operators}

Considering a six-point comb channel as in Figure~\ref{fig:Comb-six} with the conventions~\eqref{eq:prefactor_cr_conventions}, the two Casimir operators at the extremes of the comb acquire a form that is reminiscent of the five-point ones
\begin{gather}
    \widetilde{\mathcal{D}}_{12}=z_1\left[(1-z_1)z_1\partial_{z_1}^2-z_1 z_2 \partial_{z_1}\partial_{z_2}-(1-a-h_3) z_1 \partial_{z_1} + a z_2 \partial_{z_2} -a h_3\right], \label{eq:D12_sixpt_comb}\\
    \widetilde{\mathcal{D}}_{56}=z_3\left[(1-z_3)z_3\partial_{z_3}^2-z_2 z_3 \partial_{z_2}\partial_{z_3}-(1-\tilde{a}-h_4) z_3 \partial_{z_3} + \tilde{a} z_2 \partial_{z_2} -\tilde{a} h_4\right],
\end{gather}
while the Casimir operator at the middle leg is given by
\begin{multline}
    \widetilde{\mathcal{D}}_{123}=z_2\Bigl[(1-z_2)z_2\partial_{z_2}^2-z_1 z_2 \partial_{z_1}\partial_{z_2}-z_1 z_3 \partial_{z_1}\partial_{z_3}-z_2 z_3 \partial_{z_2}\partial_{z_3}-(1-h_3-h_4) z_2 \partial_{z_2} \\
    + h_3 z_3 \partial_{z_3}+ h_4 z_1 \partial_{z_1} -h_3 h_4\Bigr].
    \label{eq:sixptCas123}
\end{multline}
Note how only this last operator involves all three six-point cross-ratios and is $\mathbb{Z}_2$ invariant, while the other two are exchanged under the transformation $\mathcal{P}_6$ that exchanges $z_1\leftrightarrow z_3$, $a\leftrightarrow \tilde{a}$, and $h_3\leftrightarrow h_4$. 


\subsection{Hamiltonian interpretation for \texorpdfstring{$N=6$}{N=6} comb-channel blocks}\label{SecHamComb}

We now aim to interpret the Casimir equations for the blocks discussed in Section~\ref{ssec:Casimir_expr} as some quantum-mechanical eigenvalue equations in curved space, possibly in the presence of a background electromagnetic field.

The main idea is to see if one can construct,  starting from the Casimir operators, a quantum-mechanical Hamiltonian of the form
\begin{equation}\label{EqH}
    \mathcal{H}=-g^{ij}(\nabla_i-\iu qA_i)(\partial_j-\iu qA_j) +qV\,,
\end{equation}
where both gravity and electromagnetism are non-dynamical, and the space indices $i$ and $j$ run over the full set of cross-ratios $z_r$. Here $g^{ij}$ is the inverse metric associated with the curved space---with $\nabla_iv_j=\partial_iv_j-\Gamma_{ij}^kv_k$ the usual covariant derivative written with the help of the Christoffel symbols $\Gamma_{ij}^k$---and $A_\mu=(A_0=-V,A_i)$ is the electromagnetic background potential.\footnote{From the time-dependent Schr\"odinger equation, 
\begin{equation*}
    \iu \partial_t=\mathcal{H}=-g^{ij}(\nabla_i-\iu qA_i)(\partial_j-\iu qA_j) +qV\,,
\end{equation*}
it is straightforward to conclude from the electromagnetic covariant derivative $\partial_0-\iu qA_0=\partial_t-\iu qA_0$ that $A_0=-V$.  We note here that the time $t$ appearing in this time-dependent Schr\"odinger equation is a spurious variable that we merely used here for interpretation, as it does not mix with the space described by the inverse metric $g^{ij}$.  This time $t$ won't re-appear in the remainder of the paper.} Once a candidate Hamiltonian is determined, matching it with the form~\eqref{EqH} provides a fixed curved space and electromagnetic background field that dictate the dynamics of a quantum particle.

It is important to note that, while a priori any linear combination of the available Casimir operators could be chosen to play the role of the Hamiltonian, not all of them are suitable since the Hamiltonian~\eqref{EqH} should span the full set of ``coordinates'' $z_r$ on the curved space, \textit{i.e.}\ contain all the degrees of freedom of the system.  This makes the six-point comb-channel case a very natural one to start with, as there is a clear candidate Hamiltonian to choose from the Casimirs, namely one proportional to the middle-leg Casimir $\widetilde{\mathcal{D}}_{123}$ presented in~\eqref{eq:sixptCas123}, which is the only operator that depends on all of the three coordinates. Since the Casimirs commute, the remaining two Casimirs of the six-point function should be seen from the quantum-mechanical point of view as conserved charges with respect to the chosen Hamiltonian.

To simplify the Hamiltonian as much as possible, we will extract a simple monomial factor from the conformal blocks $\tilde{\psi}_{\{\mathbbm{h}_r\}}^{(6)}$, operation that can be interpreted as a redefinition of the conformal prefactor\footnote{We do not use this redefined prefactor from the start because our initial choice of $\Omega_N^{\{h_i\}}$ leads to the simplest behavior in the OPE limits $z_r\rightarrow 0 $, with power-law behavior that only depends on the exchanged conformal dimension: $z_r^{\mathbbm{h}_r}$.} $\Omega_6^{\{h_i\}}$. This is implemented at the level of differential operators as
\begin{equation}
    \mathcal{D}_{1\dots m}\psi_{\{\mathbbm{h}_r\}}^{(6)}=\frac{1}{\omega_6^{\{h_i\}}}\widetilde{\mathcal{D}}_{1\dots m}\omega_6^{\{h_i\}}\psi_{\{\mathbbm{h}_r\}}^{(6)}=\frac{1}{\omega_6^{\{h_i\}}}\widetilde{\mathcal{D}}_{1\dots m}\tilde{\psi}_{\{\mathbbm{h}_r\}}^{(6)}\,.
\label{def:DDtilde}
\end{equation}
We will then start our analysis from the six-point comb-channel blocks with the following choice of Hamiltonian and rescaling of prefactor:
\begin{equation}
    \mathcal{H}=-\mathcal{D}_{123}\,,\qquad\omega_6^{\{h_i\}}=z_1^{h_3}z_3^{h_4}\,,
    \label{eq:Ham6Comb_and_pref}
\end{equation}
where the minus sign in the Hamiltonian will be justified later when discussing its spectrum. This will lead to the identification of six-point blocks as some type of free-particle wavefunctions in AdS$_3$ in the next subsection, and in Section~\ref{sect:plane_waves_CB} we will work out their solution theory. We will then proceed in Section~\ref{ssec:OPE-limits-5-and-4} to analyze the five- and four-point cases as limit cases of the six-point free wavefunctions since these cases can be reached by performing OPE limits. The results of these subsections will pave the way for the general topology analysis of Section~\ref{SecGenTop}.


\subsubsection{The six-point Hamiltonian and free \texorpdfstring{AdS${}_3$}{AdS3}}
\label{ssec:six-pt-Hamiltonian}

Our starting point, as prescribed in~\eqref{eq:Ham6Comb_and_pref}, is the Hamiltonian $\mathcal{H} = -{\cal D}_{123}$ in the ``coordinates'' $z_1, z_2, z_3$, obtained by conjugating the Casimir $\widetilde{{\cal D}}_{123}$ given in~\eqref{eq:sixptCas123} with $\omega_6^{\{h_i\}}=z_1^{h_3}z_3^{h_4}$ as in~\eqref{def:DDtilde}.  This choice of $\omega_6^{\{h_i\}}$ leads us to focus on the function
\begin{equation}\label{Eqomega6}
    \psi_{\{\mathbbm{h}_1,\mathbbm{h}_2,\mathbbm{h}_3\}} (z_1,z_2,z_3)=\left(z_1^{-h_3}z_3^{-h_4}\right) \tilde{\psi}_{\{\mathbbm{h}_1,\mathbbm{h}_2,\mathbbm{h}_3\}}(z_1,z_2,z_3) \,,
\end{equation}
which constitutes an eigenfunction of the rescaled Casimir
\begin{equation}
\mathcal{D}_{123}=(1-z_2)z_2^2\partial_{z_2}^2-z_1z_2^2\partial_{z_1}\partial_{z_2}-z_3z_2^2\partial_{z_2}\partial_{z_3}-z_1z_2z_3\partial_{z_1}\partial_{z_3}-z_2^2\partial_{z_2}\,,
\label{eq:simpl-sixpt-D123}
\end{equation}
corresponding up to a sign to our Hamiltonian, and of two remaining conserved charges taking the form
\begin{equation}
\begin{gathered}
\mathcal{D}_{12}\!=\!(1-z_1)z_1^2\partial_{z_1}^2-z_1^2z_2\partial_{z_1}\partial_{z_2}\!+[2h_3-(1-a+h_3)z_1]z_1\partial_{z_1}\!-(h_3-a)z_1z_2\partial_{z_2}\!+h_3(h_3-1),\\
\mathcal{D}_{56}\!=\!(1-z_3)z_3^2\partial_{z_3}^2-z_3^2z_2\partial_{z_3}\partial_{z_2}\!+[2h_4-(1-\tilde{a}+h_4)z_3]z_3\partial_{z_3}\!-(h_4-\tilde{a})z_3z_2\partial_{z_2}\!+h_4(h_4-1).
\label{eq:simpl-sixpt-D1256}
\end{gathered}
\end{equation}

Comparing~\eqref{eq:simpl-sixpt-D123} with~\eqref{EqH}, we can identify the inverse metric $g^{ij}$ and the metric $g_{ij}$ to be
\begin{equation}
    g^{ij}=\left(
\begin{array}{ccc}
 0 & -\frac{1}{2} z_1 z_2^2 & -\frac{1}{2} z_1 z_2 z_3 \\
 -\frac{1}{2} z_1 z_2^2 & \left(1-z_2\right) z_2^2 & -\frac{1}{2} z_2^2 z_3 \\
 -\frac{1}{2} z_1 z_2 z_3 & -\frac{1}{2} z_2^2 z_3 & 0 \\
\end{array}
\right)\,, \qquad g_{ij}=\left(
\begin{array}{ccc}
 \frac{1}{z_1^2} & -\frac{1}{z_1 z_2} & -\frac{2-z_2}{z_1 z_2 z_3} \\
 -\frac{1}{z_1 z_2} & \frac{1}{z_2^2} & -\frac{1}{z_2 z_3} \\
 -\frac{2-z_2}{z_1 z_2 z_3} & -\frac{1}{z_2 z_3} & \frac{1}{z_3^2} \\
\end{array}
\right),
\label{eq:metric_z_6pt}
\end{equation}
while the electromagnetic potential $A_\mu=0$ is absent.
Here the indices $i,j=1,2,3$ correspond to $z_1,z_2,z_3$ directions, respectively.  Starting from these, we can derive the Riemann and Ricci tensors, as well as the scalar curvature $R$,
\begin{equation}
    R_{ijk\ell}=-\frac{1}{4}\left(g_{ik}g_{j\ell}-g_{jk}g_{i\ell}\right), \qquad R_{ij}=-\frac{1}{2}g_{ij}\,, \qquad R=-\frac{3}{2}\,.
\end{equation}
From the results above and the signature of the metric, we identify the $z_i$-space as the maximally-symmetric, negatively-curved AdS$_{3}$ space, with AdS radius $\alpha=2$, and vanishing background electromagnetic field, as stated above.

The free Hamiltonian and the conserved charges can be written elegantly in terms of Killing vectors, which can be determined from the metric and the Killing equations to be
\begin{equation}
\begin{gathered}
\chi_1=-\frac{1}{2} \left(1+z_1^2\right) \partial _{z_1}-\frac{z_2 \left(z_1^2+z_2-1\right) \partial _{z_2}}{2 z_1}-\frac{z_2 z_3 \partial _{z_3}}{2 z_1}\,,\\
\chi_2=\frac{1}{2} \left(z_1^2-1\right) \partial _{z_1}+\frac{\left(z_1^2-z_2+1\right) z_2 \partial _{z_2}}{2 z_1}-\frac{z_2 z_3 \partial _{z_3}}{2 z_1}\,,\\
\chi_3=z_1\partial_{z_1},\\
\chi_4=-\frac{1}{2} \left(1+z_3^2\right) \partial _{z_3}-\frac{z_1 z_2 \partial _{z_1}}{2 z_3}-\frac{z_2 \left(z_3^2+z_2-1\right) \partial _{z_2}}{2 z_3}\,,\\
\chi_5=-\frac{z_1 z_2 \partial _{z_1}}{2 z_3}+\frac{z_2 \left(z_3^2-z_2+1\right) \partial _{z_2}}{2 z_3}+\frac{1}{2} \left(z_3^2-1\right) \partial _{z_3}\,,\\
\chi_6=z_3\partial_{z_3}\,.
\end{gathered}
\label{eq:six-point-Killing-z}
\end{equation}
Their commutation relations are those of two copies of $\mathfrak{so}(2,1)$:
\begin{equation}
\begin{gathered}
{}[\chi_1,\chi_2]=-\chi_3\,,\qquad[\chi_2,\chi_3]=\chi_1\,,\qquad[\chi_3,\chi_1]=-\chi_2\,,\\
[\chi_4,\chi_5]=-\chi_6\,,\qquad[\chi_5,\chi_6]=\chi_4\,,\qquad[\chi_6,\chi_4]=-\chi_5\,.
\end{gathered}
\label{eq:six-point-Killing-comm-rel}
\end{equation}
and the Casimirs~\eqref{eq:simpl-sixpt-D123}-\eqref{eq:simpl-sixpt-D1256} take the form
\begin{equation}
\begin{gathered}
    \mathcal{D}_{123}=-\chi_1^2+\chi_2^2+\chi_3^2=-\chi_4^2+\chi_5^2+\chi_6^2\,,\\
    \mathcal{D}_{12}=\frac{1}{2}\left\{\chi_1,\chi_3 \right\}-\frac{1}{2}\left\{\chi_2,\chi_3 \right\}+\chi_3^2+\left(\frac12+a- h_3\right)\left(\chi_2-\chi_1\right)+(2h_3-1)\chi_3+h_3(h_3-1)\,,\\
    \mathcal{D}_{56}=\frac{1}{2}\left\{\chi_4,\chi_6 \right\}-\frac{1}{2}\left\{\chi_5,\chi_6 \right\}+\chi_6^2+\left(\frac12+\tilde{a}-h_4\right)\left(\chi_5-\chi_4\right)+(2h_4-1)\chi_6+h_4(h_4-1)\,.
    \label{eq:six-point-Casimirs-Killing}
\end{gathered}
\end{equation}
The free nature of the Hamiltonian ${\cal H} = -{\cal D}_{123}$ is manifest above.

\smallskip
It is instructive to change coordinates as follows,
\begin{equation}
    \rho=\frac{2 \sqrt{2}\sqrt{z_1 z_2 z_3}}{z_1+z_2+z_3-z_1 z_3 -1}\,, \quad t= \frac{\sqrt{2}\left(-z_1 -z_3+2 z_1 z_3\right)}{1-z_1-z_2-z_3+z_1 z_3}, \quad y=\frac{\sqrt{2} \left(z_1-z_3\right)}{1-z_1-z_2-z_3+z_1 z_3}\,.
   \label{eq:PoincareSix} 
\end{equation}
In these new coordinates, we end up with the following Hamiltonian:
\begin{equation}
    \mathcal{D}_{123}= \frac{\rho^2}{4}\left[\partial_\rho^2-\partial_t^2+\partial_y^2 \right]-\frac{\rho}{4} \partial_\rho\,,
\end{equation}
corresponding to the AdS$_3$ metric in Poincaré coordinates
\begin{equation}
    \dd s^2= \frac{4}{\rho^2}\left(\dd \rho^2-\dd t^2+ \dd y^2 \right),
\end{equation}
where we identify $\rho$ as the AdS depth coordinate with the AdS boundary at $\rho=0$.  

\smallskip
It is also illuminating to provide an embedding space description to gain further insight into the geometric and algebraic properties of the Hamiltonian. 
Define embedding space coordinates $X_A \in \mathbb{R}^{2,2}$ as follows
\begin{equation}
    X_1 = \frac{-1+z_2+z_1z_3}{\sqrt{z_1z_2z_3}}\,, \qquad X_2 = \frac{z_1+z_3}{\sqrt{z_1z_2z_3}}\,, \qquad X_3 = \frac{-z_1+z_3}{\sqrt{z_1z_2z_3}}\,, \qquad
    X_4 = \frac{1-z_2+z_1z_3}{ \sqrt{z_1z_2z_3}}\,,
\label{eq:EmbeddingSix}
\end{equation}
such that they are constrained to lie on the quadric $-X_1^{2}-X_2^{2}+X_3^{2}+ X_4^{2}=-4$ defining the AdS$_3$ manifold of radius $\alpha=2$.
The inverse transformation takes the form
\begin{equation}
    z_1  = \frac{X_1 + X_4}{X_2+X_3}\,, \qquad
    z_2 = \frac{4}{(X_2+X_3)(X_2-X_3)}\,, \qquad
    z_3 = \frac{X_1+X_4}{X_2-X_3}\,.
\end{equation}
This allows a direct comparison with the natural Killing vectors one can define in embedding space, corresponding to\footnote{The indices $A,B,C,D$ run from $1$ to $4$.}
\begin{equation}
    J_{AB} := X_A \frac{\partial}{\partial {X^B}} - X_B \frac{\partial}{\partial{X^A}}\,,
    \label{def:JAB}
\end{equation}
where we raise or lower indices using the flat space metric $g_{AB} = {\rm diag}(-,-,+,+)$. In fact, these
are related to the Killing vectors of~\eqref{eq:six-point-Killing-z} via
\begin{equation} 
\begin{gathered} 
\chi_1 = -\frac{J_{12}+J_{34}}{2} \,, \qquad 
\chi_2 = -\frac{J_{13}+J_{24}}{2}\,, \qquad   
\chi_3 = \frac{J_{14}-J_{23}}{2}\,, \\ 
\chi_4 = -\frac{J_{12}-J_{34}}{2}\,, \qquad  
\chi_5 = \frac{J_{13}-J_{24}}{2}\,, \qquad
\chi_6 = \frac{J_{14}+J_{23}}{2}\,.
\end{gathered}
\label{eq:six-point-Killing-emb-simple}
\end{equation}
The commutation relations~\eqref{eq:six-point-Killing-comm-rel} get repackaged as the $\mathfrak{so}(2,2)$ algebra
\begin{equation}
    [J_{AB}, J_{CD}] = g_{BC}J_{AD} - g_{AC}J_{BD}-g_{BD}J_{AC}+g_{AD}J_{BC}\,,
\end{equation}
and the Hamiltonian takes a particularly pleasing form as the quadratic Casimir
\begin{equation}
    {\cal H} = -{\cal D}_{123} = -\frac{1}{2}\left(-\chi_1^2+\chi_2^2+\chi_3^2-\chi_4^2+\chi_5^2+\chi_6^2\right) = \frac{1}{8} J_{AB} J^{AB}\,.
\end{equation}

We have thus established a perhaps unexpected relation between six-point conformal blocks in the comb channel and free-particle wavefunctions in AdS$_3$. It is important to make the following remarks:
\begin{itemize}
    \item The relation we found does not concern at all the AdS/CFT correspondence. The AdS$_3$ space that we constructed is of two dimensions higher than the CFT$_1$ we are focusing on, and the relation is established at the level of \emph{cross-ratio space} as opposed to physical space.
    \item The two conserved charges in~\eqref{eq:six-point-Casimirs-Killing} look rather unfamiliar from the perspective of AdS physics. The diagonalization of these indicates that the eigenstates that are relevant for conformal blocks are superpositions of eigenstates of the Cartan generators $\chi_3$, $\chi_6$, as we will see explicitly in the next subsection, but it is unclear whether these superpositions can correspond to AdS configurations that are relevant for other phenomena or analyses. 
    \item For the quantum-mechanical point of view to be self-consistent, it is necessary to define inner products for which the Hamiltonians~\eqref{EqH} are self-adjoint operators.  In general, the natural inner product is
\begin{equation}\label{EqIP}
    \langle\phi,\chi\rangle=\int\left[\prod_{r=1}^{N-3}dz_r\right]\sqrt{|\text{det}\,g|}\,\phi^*(z_r)\chi(z_r)\,,
\end{equation}
where $\phi$ and $\chi$ are functions that satisfy appropriate boundary conditions to ensure that surface terms generated from integration by parts vanish.  Clearly the corresponding weight function is related to the metric by $\sqrt{|\text{det}\,g|}$.

With the inner product~\eqref{EqIP} in hand, the Casimirs will be self-adjoint operators if they satisfy
\begin{equation}
    \langle\phi,\mathcal{D}_{1\cdots m}\chi\rangle=\langle\mathcal{D}_{1\cdots m}\phi,\chi\rangle\,,
\end{equation}
which is true for the cases studied above as long as the conformal dimensions satisfy $h_i^*=1-h_i$, or in other words
\begin{equation}
    h_i=\frac{1}{2}+\iu \,\lambda_i\,,
\end{equation}
with $\lambda_i\in\mathbbm{R}$, in agreement with the principal series representation. In light of this, it is now simple to see the reason for the minus sign in the definition of our Hamiltonian in~\eqref{eq:Ham6Comb_and_pref}: its eigenvalues are this way positive-definite for principal series representations
\begin{equation}
    \frac{1}{\psi_{\mathbbm{h}_2}}\mathcal{H}\psi_{\mathbbm{h}_2}=-\mathbbm{h}_2(\mathbbm{h}_2-1)=\lambda_2^2+\frac{1}{4}\,.
\end{equation}
\end{itemize}

\subsubsection{\texorpdfstring{AdS$_3$}{AdS3} wave modes and six-point conformal blocks}
\label{sect:plane_waves_CB}

Having established that the six-point conformal blocks are eigenfunctions of a free Hamiltonian in AdS$_3$ and two unconventional conserved charges, we now aim to obtain the explicit form of the six-point conformal blocks from the algebra of the AdS$_3$ Killing vectors. 

To this end, we will use the Cartan subalgebra $\{\chi_3, \chi_6\}$ that commutes with the Hamiltonian ${\cal H} = -{\cal D}_{123}$. The simultaneous eigenstates of these operators are wave modes $\varphi_{p_1,p_2,p_3}$ that satisfy
\begin{equation}
\chi_3\varphi_{p_1,p_2,p_3}=p_1\varphi_{p_1,p_2,p_3}\,,
\qquad
\chi_6\varphi_{p_1,p_2,p_3}=p_3\varphi_{p_1,p_2,p_3}\,,\qquad
{\cal H}\varphi_{p_1,p_2,p_3} = -p_2(p_2-1)\varphi_{p_1,p_2,p_3}\,.
\end{equation}
From the explicit form of the generators $\chi_3, \chi_6$ in~\eqref{eq:six-point-Killing-z} it is easily seen that the eigenstates take the form
$\varphi_{p_1,p_2,p_3}(z_1,z_2,z_3)=z_1^{p_1}z_3^{p_3}f_{p_1,p_2,p_3}(z_2)$.  The Hamiltonian~\eqref{eq:simpl-sixpt-D123} then acts like the four-point Casimir~\eqref{eq:Four-pt-GaussHyp-Diff-Eq} and leads to the well-known solutions in terms of hypergeometric functions~\cite{Dolan:2011dv},
\begin{equation}
    \frac{1}{z_1^{p_1}z_3^{p_3}}\mathcal{H}\varphi_{p_1,p_2,p_3}=-p_2(p_2-1)z_2^{p_2}{}_2F_1(p_1+p_2,p_3+p_2;2p_2;z_2)\,,
\end{equation}
with
\begin{equation}
    \varphi_{p_1,p_2,p_3}=z_1^{p_1}z_2^{p_2}z_3^{p_3}{}_2F_1(p_1+p_2,p_3+p_2;2p_2;z_2)\,.
    \label{eq:AdS_Plane_waves}
\end{equation}

The full six-point conformal block will be given by a linear superposition of these wave modes, with the asymptotics dictated by the OPE limit,
\begin{equation}
\psi_{\{\mathbbm{h}_1,\mathbbm{h}_2,\mathbbm{h}_3\}}=\psi_{\boldsymbol{\mathbbm{h}}}=\sum_{n_1,n_3\geq0}\tilde{\alpha}_{\boldsymbol{\mathbbm{h}}}(n_1,n_3)\varphi_{\mathbbm{h}_1-h_3+n_1,\mathbbm{h}_2,\mathbbm{h}_3-h_4+n_3},
    \label{eq:blocks_as_superposition}
\end{equation}
such that $\mathcal{H}\psi_{\boldsymbol{\mathbbm{h}}}=-\mathbbm{h}_2(\mathbbm{h}_2-1)\psi_{\boldsymbol{\mathbbm{h}}}$ is trivially satisfied. 
To determine the coefficients it is convenient to first define creation and annihilation operators
\begin{equation}
\begin{gathered}
    J_\pm^1:=\mp\chi_1+\chi_2\,,\qquad\qquad J_0^1:=\chi_3\,,\\
    J_\pm^3:=\mp\chi_4+\chi_5\,,\qquad\qquad J_0^3:=\chi_6\,,
\end{gathered}
\end{equation}
where the superscript indicates on which coordinates ($z_1$ or $z_3$) the generators act simply.
It is easy to verify that
\begin{equation}
[J_\pm^i,J_\mp^j]=\pm2\delta^{ij}J_0^i\,,\qquad\qquad [J_0^i,J_\pm^j]=\pm\delta^{ij}J_\pm^i\,, \qquad(\text{no sum on $i$})\,,
\end{equation}
with $i,j = 1,3$ and that the Hamiltonian $\mathcal{H}=-\mathcal{D}_{123}$ and extra conserved charges $Q^1:=\mathcal{D}_{12}$ and $Q^3:=\mathcal{D}_{56}$ can be rewritten as
\begin{equation}\label{EqHQi}
\begin{gathered}
    \mathcal{H}=-J_\pm^iJ_\mp^i-(J_0^i)^2\pm J_0^i\qquad(\text{no sum on $i$})\,,\\
    Q^i=-\frac{1}{2}\left\{J_+^i-J_0^i,\,J_0^i\right\}+(1/2+a_i-k_i)J_+^i+(2k_i-1)J_0^i+k_i(k_i-1)\,.
\end{gathered}
\end{equation}
Here $a_1\equiv a$, $a_3\equiv\tilde{a}$ while $k_1\equiv h_3$, $k_3\equiv h_4$.
The two equivalent forms of the Hamiltonian (for $i=1,3$) come from the two equivalent ways of writing the Casimir ${\cal D}_{123}^2$ in~\eqref{eq:six-point-Casimirs-Killing}.

Note that, since the algebras for the $J^1$ and $J^3$ operators commute with each other and the charges $Q^i$ only depend on the respective generators $J^i$, their diagonalization can be performed independently and the coefficients of the linear combination are expected to factorize as $\tilde{\alpha}_{\boldsymbol{\mathbbm{h}}}(n_1,n_3)=\alpha_1(n_1)\alpha_3(n_3)$.

Acting with the conserved charges leads to two additional constraints:
\begin{eqnarray}
    \mathbbm{h}_i(\mathbbm{h}_i-1)\psi_{\boldsymbol{\mathbbm{h}}}&=&Q^i\psi_{\boldsymbol{\mathbbm{h}}}  \nonumber \\
    &=&\sum_{n_1,n_3\geq0}\alpha_1(n_1)\alpha_3(n_3)Q^i\varphi_{\mathbbm{h}_1-h_3+n_1,\mathbbm{h}_2,\mathbbm{h}_3-h_4+n_3} \nonumber \\
    &=&\sum_{n_1,n_3\geq0}\alpha_1(n_1)\alpha_3(n_3)[-(\mathbbm{h}_i+n_i-a_i)J_+^i+(\mathbbm{h}_i-k_i+n_i)^2 \nonumber \\
    &&+(2k_i-1)(\mathbbm{h}_i-k_i+n_i)+k_i(k_i-1)]\varphi_{\mathbbm{h}_1-h_3+n_1,\mathbbm{h}_2,\mathbbm{h}_3-h_4+n_3}\,,
\end{eqnarray}
which generate the recurrence relations
\begin{eqnarray}
    \mathbbm{h}_i(\mathbbm{h}_i-1)\alpha_i(n_i)&=&[(\mathbbm{h}_i-k_i+n_i)^2+(2k_i-1)(\mathbbm{h}_i-k_i+n_i)+k_i(k_i-1)]\alpha_i(n_i)\nonumber\\
    &&-(\mathbbm{h}_i+n_i-1-a_i)b_{+, \mathbbm{h}_i-k_i+n_i-1,\mathbbm{h}_2}^i\alpha_i(n_i-1),
\end{eqnarray}
where we have defined\footnote{With abuse of notation we write $\{p_1,p_2,p_3\}\pm\boldsymbol{e}_i$ in the subscript of $\varphi$ to denote  $\{p_1\pm 1,p_2,p_3\}$ or $\{p_1,p_2,p_3\pm 1\}$ depending on whether $i=1$ or $3$, respectively.}
\begin{equation}\label{EqJalpha}
J_\pm^i\varphi_{p_1,p_2,p_3}=b_{\pm,p_i,p_2}^i\varphi_{\boldsymbol{p}\pm\boldsymbol{e}_i}\,.
\end{equation}
Simplifying the recurrence relation, we get
\begin{eqnarray}\label{Eqahn}
    \alpha_i(n_i)&=&\frac{(\mathbbm{h}_i+n_i-1-a_i)b^i_{+,\mathbbm{h}_i-k_i+n_i-1,\mathbbm{h}_2}}{(2\mathbbm{h}_i+n_i-1)n_i}\alpha_i(n_i-1)\nonumber\\
    &=&\frac{(\mathbbm{h}_i-a_i)_{n_i} b_{+,\mathbbm{h}_i-k_i+n_i,\mathbbm{h}_2}^i\cdots b_{+,\mathbbm{h}_i-k_i,\mathbbm{h}_2}^i}{(2\mathbbm{h}_i)_{n_i}n_i!}\alpha_i(0)\,,
\end{eqnarray}
for the coefficients of the conformal blocks. From the OPE limit asymptotics we have $\alpha_i(0)=1$. Thus we only need to figure out the proportionality constants $b^i_{+,p_i,p_2}$ from the algebra of operators \eqref{EqJalpha}.

To this end, one can proceed algebraically as follows:  Let us define $|\boldsymbol{p}\rangle$ for arbitrary representations such that\begin{equation}
\mathcal{H}|\boldsymbol{p}\rangle=-p_2(p_2-1)|\boldsymbol{p}\rangle,\qquad\qquad J_0^i|\boldsymbol{p}\rangle=p_i|\boldsymbol{p}\rangle.
\end{equation}
Thus here we have $\langle z_1,z_2,z_3|\boldsymbol{p}\rangle=\varphi_{\boldsymbol{p}}(z_1,z_2,z_3)$ and the algebra implies
\begin{equation}
J_\pm^i|\boldsymbol{p}\rangle=b_{\pm,\boldsymbol{p}}^i|\boldsymbol{p}\pm\boldsymbol{e}_i\rangle,
\end{equation}
but $(J_\pm^i)^\dagger=-J_\pm^i$, contrary to the usual $\mathfrak{su}(2)$ algebra derivation where $J_\pm^\dagger=-J_\mp$ (for anti-hermitian operators, like our Killing vectors).  Thus unitarity cannot be used as usual to figure out $b_{\pm,\boldsymbol{p}}^i$ completely.  Indeed, since the Hamiltonian can be rewritten as \eqref{EqHQi}, which implies
\begin{equation}
    b_{\pm,\boldsymbol{p}}^ib_{\mp,\boldsymbol{p}\pm\boldsymbol{e}_i}^i=p_2(p_2-1)-p_i(p_i\pm1)=(p_2\pm p_i)(p_2\mp p_i-1),
\end{equation}
a general solution can be found of the form
\begin{equation}\label{Eqalpha}
    b_{\pm,\boldsymbol{p}}^i=(p_2\pm p_i)^{1-k}(p_2\mp p_i-1)^k,
\end{equation}
for all $k$ [with $k=1/2$ the usual $\mathfrak{su}(2)$ solution].  At this point, one can argue that the proper solution should lead to representations that are lowest weight, due to the OPE limit asymptotics, which implies $k=0$ for integer-valued $p$'s (contrary to highest weight which would lead to $k=1$).

This can be verified directly by acting with the charges.  Indeed, given the AdS wave modes of~\eqref{eq:AdS_Plane_waves}, we can write down the blocks as a superposition of these, with asymptotics dictated by the OPE limit, getting~\eqref{eq:blocks_as_superposition}.
Acting on~\eqref{eq:blocks_as_superposition} with $\mathcal{D}_{12}-\mathbbm{h}_1(\mathbbm{h}_1-1)$, we get the following recurrence relation for the coefficients:
\begin{equation}
    \left(a-\mathbbm{h}_1-n_1\right) \left(-h_3+\mathbbm{h}_1+\mathbbm{h}_2+n_1\right)\alpha_1(n_1) +(n_1+1) \left(2 \mathbbm{h}_1+n_1\right)\alpha_1(n_1+1)=0\,,
\end{equation}
with an analogous recurrence relation from $\mathcal{D}_{56}$ which is the $\mathbb{Z}_2$-transformed of the latter one. These can be easily solved, and imposing $\alpha_i(0)=1$, compatibly with the OPE limit, we get the solution
\begin{equation}
    \alpha_i(n_i)=\frac{\left(\mathbbm{h}_i-a_i\right)_{n_i} \left(-k_i+\mathbbm{h}_i+\mathbbm{h}_2\right)_{n_i}}{n_i! \left(2 \mathbbm{h}_i\right)_{n_i}} \,,
    \label{eq:coefficients_six_points}
\end{equation}
in agreement with \eqref{Eqahn} and \eqref{Eqalpha} with $k=0$.
Putting everything together, we get the solution of six-point blocks as
\begin{multline}\label{EqBlock6ptComb}
    \tilde{\psi}_{\mathbbm{h}_1,\mathbbm{h}_2,\mathbbm{h}_3}=\sum_{n_1\ge0}\sum_{n_3\ge 0} \frac{\left(\mathbbm{h}_1-a\right)_{n_1} \left(-h_3+\mathbbm{h}_1+\mathbbm{h}_2\right)_{n_1}}{n_1! \left(2 \mathbbm{h}_1\right)_{n_1}} \frac{\left(\mathbbm{h}_3-\tilde{a}\right)_{n_3} \left(-h_4+\mathbbm{h}_2+\mathbbm{h}_3\right)_{n_3}}{n_3! \left(2 \mathbbm{h}_3\right)_{n_3}}\\
    \times z_1^{\mathbbm{h}_1+n_1}z_2^{\mathbbm{h}_2}z_3^{\mathbbm{h}_3+n_3}{}_2F_1\!\left[\begin{array}{c}\mathbbm{h}_1+\mathbbm{h}_2-h_3+n_1,\,\,\mathbbm{h}_2+\mathbbm{h}_3-h_4+n_3\\
    2\mathbbm{h}_2\end{array};z_2\right]\,,
\end{multline}
where we also re-included the factors of $z_1^{h_3} z_3^{h_4}$ that we had extracted. This expression can be shown to agree with (B.49) of~\cite{Buric:2021kgy} (obtained using the results of~\cite{Rosenhaus:2018zqn}) by simply series expanding the hypergeometric functions and repackaging the summations in a different way, showing explicitly the sum over AdS$_3$ wave modes.

\subsection{OPE limits to five- and four-point Hamiltonians}
\label{ssec:OPE-limits-5-and-4}

Having completed the discussion on six-point conformal blocks, we can now start to study the lower-point cases. Since these can be derived by simply taking OPE limits of the six-point case, it is obvious that these should then correspond to certain limits of free-particle wavefunctions in AdS$_3$. It will be nonetheless interesting to analyze what these limits correspond to in the AdS picture and study which types of effective Hamiltonians one recovers in the relevant lower-dimensional subspaces of AdS$_3$. 

OPE limits correspond to configurations in physical space where external positions of fields that are paired in an OPE diagram collide. In this paper, we will always work in sets of cross ratios $z_r$ that have the property that for any OPE limit $x_{i}\rightarrow x_j$, we have one unique cross-ratio $z_m$ that is sent to zero. This implies that any cross-ratio $z_r$ is uniquely associated with a single internal leg, and that our conformal blocks $\tilde{\psi}_{\boldsymbol{\mathbbm{h}}}^{(\chan_N)}$ behave in any OPE limit as a simple power law of the vanishing cross-ratio $z_m^{\mathbbm{h}_m}$---which picks up only the contribution of the primary operator exchanged in the $z_m$ internal leg---multiplied by a $(N-1)$-point conformal block $\tilde{\psi}_{\boldsymbol{\mathbbm{h}}}^{(\chan_{N-1})}$ that does not depend on $z_m$.

At the level of differential operators, one can use the OPE limits to reduce $N$-point Casimir operators to those for $M<N$ points. This can be achieved by iteratively extracting the leading asymptotics of the blocks from the differential operators and taking the $z_m\rightarrow 0$ limit as follows
\begin{equation}
    \widetilde{\mathcal{D}}_{p'\dots q'}^{(\chan_{N-1})}=\lim_{z_m\rightarrow 0}\frac{1}{z_m^{\mathbbm{h}_m}} \widetilde{\mathcal{D}}_{p\dots q}^{(\chan_{N})} z_m^{\mathbbm{h}_m}\,,
    \label{eq:general_OPE_limit}
\end{equation}
where the indices for the summation over generators in the Casimir operators are not necessarily identical.

This limit is such that, if we name $\widetilde{\mathcal{D}}_m$ the Casimir operator associated with the internal leg of $z_m$, the $\widetilde{\mathcal{D}}_m$ Casimir gets trivialized to a constant (as expected for an external operator)
\begin{equation}
    \lim_{z_m\rightarrow 0}\frac{1}{z_m^{\mathbbm{h}_m}} \widetilde{\mathcal{D}}_{m} z_m^{\mathbbm{h}_m}=\mathbbm{h}_m(\mathbbm{h}_m-1)
\end{equation}
and the remaining Casimir operators become those of the $(N-1)$-point function.

We can now move to the application of these OPE limits to discuss the five- and four-point conformal blocks.


\subsubsection{Five-point Hamiltonian: \texorpdfstring{AdS${}_2$}{AdS2} with constant background magnetic field}
\label{ssec:5-pt-Hamiltonian}

In order to reduce from six- to five-point comb-channel blocks, there are two possible OPE limits that one can consider, corresponding to the two colliding limits that can happen at the two sides of the comb, which in cross-ratio space are translated to $z_1\to 0$ and $z_3 \to 0$.

In Poincar\'{e} coordinates~\eqref{eq:PoincareSix}, these OPE limits correspond to approaching the AdS$_3$ boundary $\rho \to 0$ with respectively $y-t \to 0$ or $y+t \to 0$.\footnote{With the ratios $\rho^2/(y\mp t)$ held fixed.} 
The same picture in embedding space~\eqref{eq:EmbeddingSix}, is the approaching of the conformal boundary of AdS$_3$ at infinity, $-X_1^2 - X_2^2 + X_3^2 + X_4^2 = 0$, along the null directions:
\begin{gather} 
z_1 \to 0 \qquad  \Longrightarrow \qquad (X_1  + X_4 \to 0\,, X_2 - X_3 \to 0)\,, \\ 
z_3 \to 0 \qquad \Longrightarrow \qquad (X_1  + X_4 \to 0\,, X_2 + X_3 \to 0)\,.
\end{gather}
We will soon see that these limits are such that the space that is recovered at the conformal boundary is actually AdS$_2$.

Without loss of generality due to the $\mathbb{Z}_2$ symmetry of the six-point comb channel, we will consider the reduction to five points via the $x_1\rightarrow x_2$ OPE limit, corresponding to the $z_1\rightarrow 0$ limit. 
Using the recipe of~\eqref{eq:general_OPE_limit}, the six-point Casimir equations for $\tilde{\psi}_{\mathbbm{h}_1,\mathbbm{h}_2,\mathbbm{h}_3}^{(6)}$ reduce to the five-point Casimir equations for the blocks associated with the $\expval{\mathcal{O}_{\mathbbm{h}_1}(x_2)\phi_3\phi_4\phi_5\phi_6}$ correlator, multiplied by power-law $z_1^{\mathbbm{h}_1}$ with respect to the variable that is sent to zero. 
Considering that for the six-point Hamiltonian and conserved charges we extracted a $z_1^{h_3}$ in~\eqref{Eqomega6}, we need to extract a power of $z_1^{\mathbbm{h}_1-h_3}$ in~\eqref{eq:general_OPE_limit} in order to perform the limit correctly. 
Furthermore, in order to work in the same conventions as the five-point expressions of Section~\ref{ssec:5-pt-Casimirs}, we rename fields, conformal dimensions, and insertion points for the reduced correlator such that
\begin{equation}
    \expval{\mathcal{O}_{\mathbbm{h}_1}(x_2)\phi_3\phi_4\phi_5\phi_6} \,\,\, \stackrel{\mathrm{redef.}}{\longrightarrow} \,\,\, \expval{\phi_1\phi_2\phi_3\phi_4\phi_5}
\end{equation}
and the cross ratios are shifted as $z_2,z_3 \rightarrow z_1,z_2$.

Applying this recipe to the six-point differential operators, we see that, consistently, $\mathcal{D}_{12}^{(6)}$ gets reduced to a constant for the first external field
\begin{equation}
    \lim_{z_1\rightarrow 0}z_1^{h_3-\mathbbm{h}_1} \mathcal{D}_{12}^{(6)} z_1^{\mathbbm{h}_1-h_3}=\mathbbm{h}_1(\mathbbm{h}_1-1) \,\,\, \stackrel{\mathrm{redef.}}{\longrightarrow} \,\,\, h_1(h_1-1)\,.
\end{equation}
For the non-trivial operators, the six-point Hamiltonian $\mathcal{H}=-\mathcal{D}_{123}^{(6)}$ gets reduced to $\mathcal{H}=-\mathcal{D}_{12}^{(5)}$ with
\begin{equation}\label{EqH5pts}
\mathcal{D}_{12}^{(5)}=(1-z_1)z_1^2\partial_{z_1}^2-z_1^2z_2\partial_{z_1}\partial_{z_2}-(1-a)z_1^2\partial_{z_1}+az_1z_2\partial_{z_2}\,,
\end{equation}
and the conserved charge $\mathcal{D}_{56}^{(6)}$ becomes
\begin{equation}\label{EqD45}
\mathcal{D}_{45}^{(5)}=(1-z_2)z_2^2\partial_{z_2}^2-z_1z_2^2\partial_{z_1}\partial_{z_2}+(\tilde{a}-h_3)z_1z_2\partial_{z_1}+[2h_3-(1-\tilde{a}+h_3)z_2]z_2\partial_{z_2}+h_3(h_3-1)\,.
\end{equation}
It is straightforward to check that~\eqref{EqH5pts} and~\eqref{EqD45} correspond to the operators~\eqref{eq:fiveptCas12} and~\eqref{eq:fiveptCas45} after extraction of the factor $\omega_5^{\{h_i\}}=z_2^{h_3}$.
Note that if we had taken instead the $z_3\rightarrow 0$ limit from the six-point expressions, we would have ended up in a $\mathbb{Z}_2$-reflected configuration, with Hamiltonian equal up to a sign to $\mathcal{D}_{45}^{(5)}$.

Comparing~\eqref{EqH5pts} with~\eqref{EqH}, it is simple to derive the following (inverse) metric
\begin{equation}
    g^{ij}=\left( \begin{array}{cc}
      z_1^2(1-z_1)   & -\frac{1}{2}z_1^2 z_2 \\
       -\frac{1}{2}z_1^2 z_2  & 0
    \end{array} \right), \qquad \qquad g_{ij}=\left( \begin{array}{cc}
     0   & -\frac{2}{z_1^2 z_2} \\
       -\frac{2}{z_1^2 z_2}  & -4\frac{1-z_1}{z_1^2 z_2^2}
    \end{array} \right),
\end{equation}
which leads to the following Ricci tensor and scalar curvature:
\begin{equation}
    R_{ij}=-g_{ij} \qquad \Longrightarrow \qquad R=-2\,.
\end{equation}
Since we are talking about a two-dimensional space, this implies that the space is maximally symmetric, and corresponds in particular to AdS$_2$ space with radius $\alpha=1$.

Contrary to the six-point case, there is a non-vanishing background electromagnetic field in the five-point case.  Comparing~\eqref{EqH5pts} with~\eqref{EqH} again, the background electromagnetic potential is
\begin{equation}\label{EqA5pts}
    A_\mu=(A_0,A_1,A_2)=\left(\iu a,\frac{1}{z_1},\frac{2-z_1}{z_1 z_2}\right)\,,
\end{equation}
interacting with a charged particle of charge $q=-\iu a$.\footnote{This is consistent with the requirement of self-adjointness of the Hamiltonian, which is achieved for principal series representations, see last comment of Section~\ref{ssec:six-pt-Hamiltonian}.}  Constructing the background electromagnetic tensor field from~\eqref{EqA5pts} leads to
\begin{equation}
    F_{\mu\nu}=\partial_\mu A_\nu-\partial_\nu A_\mu=\left(\begin{array}{ccc}0&0&0\\0&0&\frac{2}{z_1^2z_2}\\0&-\frac{2}{z_1^2z_2}&0\end{array}\right)\,,
\end{equation}
which corresponds to a vanishing background electric field and a constant background magnetic field $B_\perp=F_{23}/\sqrt{|\text{det}\,g|}=1$ perpendicular to the curved surface.

Hence, the five-point Hamiltonian describes a charged particle of charge $-\iu a$ moving on a curved AdS$_2$ space with radius $\alpha=1$ and a constant background magnetic field $B_\perp=1$ perpendicular to the surface.

As for the six-point case, the Casimirs can be expressed quite elegantly in terms of the Killing vectors.  From the metric and the Killing equation, the Killing vectors are
\begin{equation}
\begin{gathered}
\chi_1=\frac{z_1(1-z_1-z_2^2)}{2z_2}\,\partial_{z_1}-\frac{1+z_2^2}{2}\,\partial_{z_2}\,,\\
\chi_2=\frac{z_1(1-z_1+z_2^2)}{2z_2}\,\partial_{z_1}-\frac{1-z_2^2}{2}\,\partial_{z_2}\,,\\
\chi_3=z_2\partial_{z_2}\,,
\end{gathered}
\label{eq:five-point-Killing-z}
\end{equation}
and they satisfy the following algebra
\begin{equation}
[\chi_1,\chi_2]=-\chi_3\,,\qquad[\chi_2,\chi_3]=\chi_1\,,\qquad[\chi_3,\chi_1]=-\chi_2\,,
\end{equation}
which corresponds to one copy of $\mathfrak{so}(2,1)$, \textit{i.e.}\ half of the six-point case~\eqref{eq:six-point-Killing-comm-rel}.

In terms of the Killing vectors,~\eqref{EqH5pts} and~\eqref{EqD45} become
\begin{equation}
\begin{gathered}
\mathcal{D}_{12}^{(5)}=-\left(\chi_1+\frac{a}{2}\frac{z_1}{z_2}\right)^2+\left(\chi_2+\frac{a}{2}\frac{z_1}{z_2}\right)^2+\chi_3^2\,,\\
\mathcal{D}_{45}^{(5)}=\frac{1}{2}\{\chi_1,\chi_3\}-\frac{1}{2}\{\chi_2,\chi_3\}+\chi_3^2-(\tilde{a}-h_3+1/2)(\chi_1-\chi_2)+(2h_3-1)\chi_3+h_3(h_3-1)\,.
\end{gathered}
\end{equation}

To better understand the structure of the space, it is again instructive to change coordinates to
\begin{equation}
    z_1=\frac{2\rho}{\rho+x},\qquad z_2=\frac{1}{\rho+x}\,,
    \label{eq:to_conf_flat_five}
\end{equation}
where a conformally flat metric for AdS$_2$ is reached.  In these coordinates, the Casimirs become
\begin{equation}
\begin{gathered}
\mathcal{D}_{12}^{(5)}=\rho^2(\partial_\rho^2-\partial_x^2)-2a\rho\partial_x\,,\\
\mathcal{D}_{45}^{(5)}=\rho^2\partial_\rho^2-(1-2x)\rho\partial_\rho\partial_x-(1-x)x\partial_x^2-[1+\tilde{a}-h_3]\partial_x+(h_3-1)(h_3-2\rho \partial_\rho-2x\partial_x)\,,
\end{gathered}
\label{eq:five-pt-conf_flat_ops}
\end{equation}
where $\rho$ is identified with the AdS depth coordinate, with AdS boundary at $\rho=0$, reminiscent of the six-point case.  

Once again, we can map the above coordinates to the embedding space $\mathbb{R}^{2,1}$ with signature $(-,-,+)$ via
\begin{equation}
    X_1 = \frac{-1+z_1+z_2^2}{z_1 z_2}\,, \qquad X_2 = \frac{2-z_1}{z_1}\,, \qquad X_3 = \frac{-1+z_1-z_2^2}{z_1 z_2} \,,
\end{equation}
where AdS$_2$ is given by the quadric
\begin{equation}
    -X_1^2-X_2^2+X_3^2=-1\,.
\end{equation}
The inverse transformation is
\begin{equation}
    z_1 = \frac{2}{1+X_2}\,, \qquad z_2 = \frac{X_1-X_3}{1+X_2}\,.
\end{equation}
In embedding space the Killing vectors \eqref{eq:five-point-Killing-z} are identified with
\begin{equation}
    \chi_1 = -J_{12} \qquad \chi_2 = J_{23} \qquad \chi_3 = J_{31}\,,
\end{equation}
where again
\begin{equation}
    J_{AB} \equiv X_A \frac{\partial}{\partial {X^B}} - X_B \frac{\partial}{\partial{X^A}}\,,
    \label{def:JAB5pts}
\end{equation}
and we raise and lower indices using the metric $g_{AB} = {\rm diag}(-,-,+)$ with $A,B = 1,2,3$.

To find the five-point conformal blocks, it is possible to proceed starting with wave modes as in the six-point case.  However, it is much simpler to implement the OPE limit on \eqref{EqBlock6ptComb}, extracting the power of the primary, resulting in
\begin{equation}\label{EqBlock5ptComb}
    \tilde{\psi}_{\mathbbm{h}_1,\mathbbm{h}_2}=\sum_{n_2\ge 0} \frac{\left(\mathbbm{h}_2-\tilde{a}\right)_{n_2} \left(-h_3+\mathbbm{h}_1+\mathbbm{h}_2\right)_{n_2}}{n_2! \left(2 \mathbbm{h}_2\right)_{n_2}}z_1^{\mathbbm{h}_1}z_2^{\mathbbm{h}_2+n_2}{}_2F_1\!\left[\begin{array}{c}\mathbbm{h}_1-a,\,\,\mathbbm{h}_1+\mathbbm{h}_2-h_3+n_2\\
    2\mathbbm{h}_1\end{array};z_1\right]\,
\end{equation}
after reintroducing the $z_2^{h_3}$ factor. It is easy to verify that the expression obtained here agrees with equation (2.10) of~\cite{Rosenhaus:2018zqn} once one expands the $_2F_1$ as a series around $z_1=0$.


\subsubsection{Four-point Hamiltonian: P\"oschl-Teller potential}
\label{ssec:4-pt-Hamiltonian}

Having obtained an AdS$_2$ Hamiltonian in the previous subsection, we can now study what happens if we perform one final limit to reduce to the four-point case. Since our five-point Hamiltonian is associated with the leg of $z_1$, it only makes sense to consider the $z_2\to 0$ limit.\footnote{If one were to look for an interpretation of the $z_1\to 0$ limit, one could consider the $\mathbb{Z}_2$-reflected case in which the Hamiltonian is $\mathcal{H}=-\mathcal{D}_{45}$, so our analysis of the $z_2\to 0$ case is without loss of generality.} In the conformally-flat coordinates of~\eqref{eq:to_conf_flat_five}, this limit corresponds to taking $\rho+x\to\pm\infty$  keeping the ratio $\rho/x$ fixed, while in embedding-space coordinates this OPE limit corresponds to the section of AdS$_2$ where $X_1 - X_3 \to 0$ while $X_2$ remains free. Both of these limits are not so intuitive, so to better understand the causal picture behind this limit we change coordinates to the following
\begin{equation}
    z_1= 1-e^{-2 \sigma}\,, \qquad z_2 = e^{-\sigma-t}\,,
\end{equation}
for which the five-point Hamiltonian acquires a different conformally flat form
\begin{equation}
    \mathcal{D}_{12}^{(5)} \psi(\sigma,t)= \sinh^2\!\sigma \left( \partial_\sigma^2-\partial_t^2\right) \psi + 2 a \sinh \sigma (\sinh \sigma \, \partial_\sigma - \cosh \sigma \, \partial_t)\psi\,,
    \label{eq:fivept_Cas_st}
\end{equation}
with (inverse) metric in $(\sigma,t)$ coordinates
\begin{equation}
    g^{ij}=\sinh^2\! \sigma\left( \begin{array}{cc}
      1   & 0 \\
       0  & -1
    \end{array} \right) \qquad \qquad g_{ij}=\frac{1}{\sinh^2\! \sigma}\left( \begin{array}{cc}
      1   & 0 \\
       0  & -1
    \end{array} \right).
\end{equation}
In these coordinates, it is straightforward to see that taking the $z_2\to 0$ OPE limit corresponds to following a time-like curve with fixed $\sigma$ all the way to $t\rightarrow +\infty$.

In performing the $t\to +\infty$ limit for the conformal blocks we extract, together with the usual leading behavior under the OPE limit, a convenient prefactor in the following way:
\begin{equation}
    \psi^{(5)}(\sigma,t)\sim \sqrt{1-e^{-2 \sigma}} e^{-a \sigma+b t+\frac{\sigma}{2}} \psi^{(4)}(\sigma)\,, \qquad \text{with} \quad b=h_3-\mathbbm{h}_2.
\end{equation}
Plugging this leading behavior in the five-point Hamiltonian $\mathcal{D}_{12}^{(5)}$ and changing variables to
\begin{equation}
    \xi=2 \log\left(\frac{1+e^{-\sigma}}{\sqrt{1-e^{-2\sigma}}}\right)
\end{equation}
we get the following operator acting on $\psi^{(4)}(\xi)$:
\begin{equation}
    \mathcal{H}=-\partial_\xi^2+\frac{(a-b)^2-\frac{1}{4}}{\sinh^2\xi}+\frac{ab}{\sinh^2\frac{\xi}{2}}+\frac{1}{4}\,.
\end{equation}
Following the time-like curve has then left us with just one spatial dimension, on which our initial Hamiltonian reduces to a quantum-mechanical kinetic term plus the well-known integrable P\"oschl-Teller potential~\cite{Poschl:1933zz}; compare also with~\cite[equation~11]{Isachenkov:2016gim} with a sign difference due to our (real) variable $\xi$ differing from their variable $x$ by the constant $\iu \pi$. We have this way recovered the relationship between one-dimensional four-point conformal blocks and the P\"oschl-Teller Hamiltonian, initially discovered in~\cite{Isachenkov:2016gim}, and we also underlined in passing how this integrable model arises in the context of free theory in AdS$_3$ (cf. \cite{ DaRocha:2005pr,Correa:2008bi,Lagogiannis:2011st}).

\subsection{Reduction via replacement with identity operator: identity limits}
\label{ssec:identity_reduction}

While the OPE limits we analyzed in Section~\ref{ssec:OPE-limits-5-and-4} give perhaps the most intuitive description of how the five- and four-point blocks can be recovered from limits of AdS$_3$ free-particle wavefunctions, there is another procedure one can follow to get to the same results, and which will turn very useful in the context of $(N>6)$-point conformal blocks. We call this procedure an ``identity limit'', consisting of replacing one of the external fields $\phi_k(x_k)$ in the $N$-point correlator with an identity operator $\mathds{1}(x_k)$. This identity insertion should play no role in the correlator and dependence on its insertion point $x_k$ should simply drop out, reducing to a correlator with $N-1$ external legs. At the level of the block expansion, this is implemented by setting the $k$-th conformal dimension $h_k=0$ to vanish, and imposing equal fields in the two legs that are attached to the $k$-th external leg of the OPE diagram. Naming $\bar{\nu}$ the vertex where the $k$-th leg is inserted, we have for the leg with the identity
\begin{equation}
    \mathcal{D}_{\bar{\nu}_{(1)}}^{(\chan_N)}\psi^{(\chan_{(N-1)+\mathds{1}})}\equiv \mathcal{D}_{k}^{(\chan_N)}\psi^{(\chan_{(N-1)+\mathds{1}})}=h_k(h_k-1)\psi^{(\chan_{(N-1)+\mathds{1}})}=0
\end{equation}
and the equality between the Casimirs associated with the other two legs of the vertex $\bar{\nu}$
\begin{equation}
    \mathcal{D}_{\bar\nu_{(2)}}^{(\chan_N)}\psi^{(\chan_{(N-1)+\mathds{1}})} \Bigl|_{h_k=0}=\mathcal{D}_{\bar\nu_{(3)}}^{(\chan_N)}\psi^{(\chan_{(N-1)+\mathds{1}})}\Bigl|_{h_k=0}\,.
\end{equation}

Focusing on the six-point comb-channel case from the AdS perspective, since the parameter $h_k$ does not appear in the Hamiltonian, this limit corresponds to setting one of the conserved charges equal to the Hamiltonian and thus represents a specific sub-case of the AdS$_3$ free-particle wavefunctions we described in Section~\ref{SecHamComb}.
With every limit, the AdS waves get constrained to a codimension 1 surface in cross-ratio space up to a prefactor, thus implementing the reduction to a lower-point block.

\subsubsection{Example: identity limit from six points to five}
\label{subsec:example-identity-lim-6pt}
To provide a concrete example, let us consider the reduction from the six-point blocks for the $\expval{\phi_1\phi_2\phi_3\phi_4\phi_5\phi_6}$ correlator to the five-point blocks for $\expval{\phi_1\phi_2\phi_3\phi_5\phi_6}$ via the identity limit on the fourth external leg. Setting $h_4=0$ and requiring the equality $\mathcal{D}_{123}=\mathcal{D}_{56}$, we expect the dependence on the coordinate $x_4$ to be present only through the prefactor $\Omega_6$ that was chosen for the six-point function. If we therefore divide the blocks by 
\begin{equation}
\tilde{\omega}_6=\frac{\Omega_6}{\Omega_{\expval{\phi_1\phi_2\phi_3\phi_5\phi_6}}}\biggl|_{h_4=0} = (1-z_2)^{-h_3}(1-z_3)^{h_6-h_5}\,,  
\end{equation}
the leftover expression should coincide with the expected five-point block. Explicitly, introducing the five-point cross-ratios for the $\expval{\phi_1\phi_2\phi_3\phi_5\phi_6}$ correlator
\begin{equation}
    \zeta_1=\frac{z_1}{1-z_2}\,, \qquad \zeta_2=\frac{z_2 z_3}{(1-z_2)(1-z_3)}\,,
\end{equation}
we have
\begin{equation}
    \tilde{\omega}_6 \widetilde{\mathcal{D}}_{12}^{(6)} \frac{\tilde{\psi}^{(5+\mathds{1})}(\zeta_1,\zeta_2,z_3)}{\tilde{\omega}_6}=\widetilde{\mathcal{D}}_{12}^{(5)} \tilde{\psi}^{(5)}(\zeta_1,\zeta_2)
\end{equation}
and the equality of the other two Casimirs imposes the constraint
\begin{multline}
     \tilde{\omega}_6 \left(\widetilde{\mathcal{D}}_{123}^{(6)}-\widetilde{\mathcal{D}}_{56}^{(6)} \right) \frac{\tilde{\psi}^{(5+\mathds{1})}(\zeta_1,\zeta_2,z_3)}{\tilde{\omega}_6}\biggl|_{h_4=0}=\\
     \Bigl[z_3^2 \left(z_3-1\right) \partial _{z_3}-\zeta _2 \left(\zeta _2+\left(\zeta _2-1\right) \left(z_3-2\right)z_3\right)\partial _{\zeta _2}-\zeta _1 \zeta _2 \left(z_3-1\right){}^2 \partial _{\zeta _1}\\
     z_3^2 \left(\tilde{a}+\zeta _2 h_3+1\right)+\zeta _2 h_3-2 \zeta _2 h_3 z_3\Bigr]\partial _{z_3}\tilde{\psi}^{(5+\mathds{1})}(\zeta_1,\zeta_2,z_3)=0\,,
\end{multline}
which enforced on the two individual Casimirs produces two identical differential operators that commute with $(1-z_3)\partial_{z_3}$:
\begin{multline}
\widetilde{\mathcal{D}}_{123}^{(6)}\tilde{\psi}^{(5+\mathds{1})}(\zeta_1,\zeta_2,z_3)=\widetilde{\mathcal{D}}_{56}^{(6)}\tilde{\psi}^{(5+\mathds{1})}(\zeta_1,\zeta_2,z_3)=
\zeta _2\Bigl[\left(1-\zeta _2\right) \zeta _2 \partial _{\zeta _2}^2-\zeta _1 \zeta _2 \partial _{\zeta _1} \partial _{\zeta _2}\\
+\left(h_3-1\right)\zeta _2 \partial _{\zeta _2}+\left(\left(1-z_3\right) \partial _{z_3}-\tilde{a}\right)\left(h_3-\zeta_1\partial_{\zeta_1}-\zeta _2 \partial_{\zeta_2}\right)\Bigr]   \tilde{\psi}^{(5+\mathds{1})}(\zeta_1,\zeta_2,z_3) \,.
\label{eq:identity-limit-Casimir}
\end{multline}
We can then look for solutions that also diagonalize $(1-z_3)\partial_{z_3}$, so that they have an overall power-law dependence $(1-z_3)^{p}$. It's simple to check from~\eqref{eq:identity-limit-Casimir} that any power law of this sort leads to Casimirs of the same form as the five-point $\widetilde{\mathcal{D}}_{45}^{(5)}$ up to a shift of the parameter $\tilde{a}$. In order to reproduce a Casimir with the same conventions as~\eqref{eq:fiveptCas45}, we then just need to consider the $p=0$ solution, as expected from the ratio of prefactors $\tilde{\omega}_6$. Imposing this, we then have
\begin{equation}
    \tilde{\omega}_6 \widetilde{\mathcal{D}}_{123}^{(6)} \frac{\tilde{\psi}^{(5+\mathds{1})}(\zeta_1,\zeta_2)}{\tilde{\omega}_6}=\tilde{\omega}_6 \widetilde{\mathcal{D}}_{56}^{(6)} \frac{\tilde{\psi}^{(5+\mathds{1})}(\zeta_1,\zeta_2)}{\tilde{\omega}_6}=\widetilde{\mathcal{D}}_{56}^{(5)} \tilde{\psi}^{(5)}(\zeta_1,\zeta_2)\,,
\end{equation}
which confirms that the six-point blocks reduce under the identity limit to
\begin{equation}
    \tilde{\psi}^{(6)}(z_1,z_2,z_3)\Bigl|_{h_4=0,\mathbbm{h}_2=\mathbbm{h}_3}=\tilde{\omega}_6\tilde{\psi}^{(5)}(\zeta_1,\zeta_2)\,.
\end{equation}
Similar types of reductions can be done from five to four points, but also in lower- and higher-point cases. While we will discuss another explicit example only in Appendix~\ref{app:11-point-limit}, these identity limits will be very useful in Section~\ref{subsect:non_6PC_diagrams} to relate any one-dimensional conformal block for any topology to AdS$_3^{\otimes m}$ free-particle wavefunctions.

\section{General topology}\label{SecGenTop}
In this section, we aim to generalize the results of the previous section to OPE channels of any topology. As we will see, any $N$-point block can be seen as (limits of) AdS$_3^{\otimes (\lceil N/3\rceil-1 )}$ free-particle wavefunctions. The strategy we will adopt to prove this statement can be structured along the following points:
\begin{itemize}
    \item We specify a general convention for the set of cross-ratios and prefactors that keeps the coupling between the various components of an OPE channel as localized as possible.
    \item We argue that for every $N=(3+3m)$-point OPE diagram that can be obtained by ``gluing'' six-point diagrams along the extremal vertices, its conformal blocks can be understood as a linear combination of AdS$_3^{\otimes m}$ wave modes of equal energy, with coefficients that can be computed explicitly.
    \item We argue how blocks for every topology that is not captured by the previous point can be seen as an explicit limit of some higher-point block that falls in the category above.
\end{itemize}
After completing these steps, we will conclude by showing the Feynman rules one can use to produce any one- or two-dimensional $N$-point conformal block in the conventions of this paper.

\subsection{Conventions for arbitrary OPE channels}
\label{subsect:Conventions_Flow}
\begin{figure}[htp]
    \centering
    \includegraphics[scale=0.75]{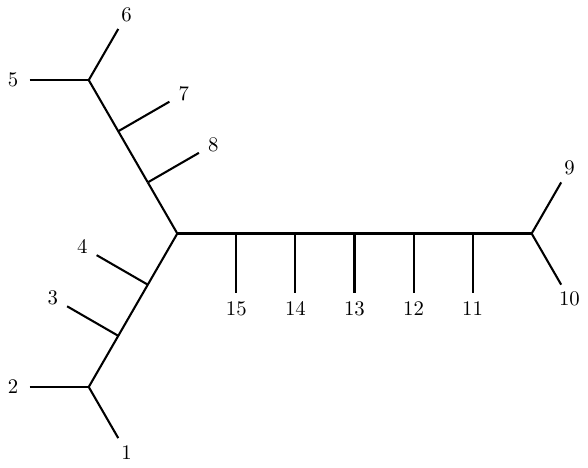}
    \caption{Example of an OPE diagram that is compatible with our convention for numbering external legs. These have to be disposed in a planar way, with teeth of any comb disposed on the same side, and such that the numbering is in ascending order when following the perimeter of the diagram in a clockwise direction starting from the first leg.}
    \label{fig:15-points}
\end{figure}

In order to have formulas that can apply to any number of points and any topology, it is important to establish convenient conventions for all of the moving parts that change when considering different channels. We will always represent the OPE channels in a planar way, where we draw all parts that have successive OPEs that involve one single external operator as comb-like structures, with all the field insertions located on the same side of the comb.
The fields within the diagram will then be labeled cyclically, as if we were to  embed the planar diagram in a circle, with external points located along the circumference. With this type of labeling, we can assume without loss of generality that $x_1>x_2>\dots>x_N$, as the channel expansion will be convergent in this domain. An example of a diagram consistent with our conventions is in Figure~\ref{fig:15-points}, where we draw a 15-point OPE channel.

To express our conventions, we will be using a one-dimensional adaptation of the ``OPE flow'' of~\cite{Fortin:2022grf}, which consists in a graphical way to encode convenient choices of conformal prefactors and cross-ratios tailored to specific OPE channels. We will review this construction to the extent that is needed for the one-dimensional blocks in analysis. 

\begin{figure}[htp]
    \centering
    \includegraphics[scale=1]{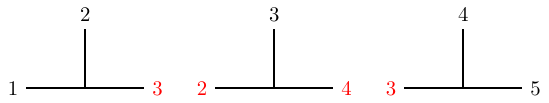}
    \caption{Example of a flow diagram for a five-point conformal block. This fully specifies a convention for around which positions OPE limits should be taken.}
    \label{fig:five-pt-flow}
\end{figure}

The starting point is to decompose the OPE diagram in a set of $N-2$ three-point vertices, each with three field insertions localized at explicit external points. We will call this the \emph{flow diagram} of the OPE channel $\chan$, and we will label it $f_{\chan}$. A given flow diagram $f_\chan$ completely fixes the conventions for the OPE channel, cross-ratios, and conformal prefactor that are associated with the conformal block $\psi^{(\chan)}$. The idea behind the flow diagram is to conventionally fix for any pair of fields that could be considered coupled in an OPE, one external position around which the expansion can be performed.  For instance, for a five-point function we can write down the decomposition as in Figure~\ref{fig:five-pt-flow}. Read from left to right, this diagram signifies that: the $\phi_1(x_1)\times \phi_2(x_2)$ OPE is to be performed around the position $x_2$, which is why this coordinate appears in the second vertex; 
at that second vertex, we can read off that if we were to take the $\mathcal{O}_1(x_2)\times \phi_3(x_3)$ OPE, then we would expand around the position $x_3$, which again motivates the presence of $x_3$ in the third vertex. But what we just described is not the only way in which the OPEs could be performed to construct this channel. One could analogously perform OPEs from right to left, or perform one OPE on the right and one OPE on the left. In all of these cases, one should end up with the same conformal blocks. The consistent labeling of Figure~\ref{fig:five-pt-flow} is meant to take into account all those possible directions in which OPEs can be performed.

\begin{figure}[htp]
    \centering
    \includegraphics[scale=0.5]{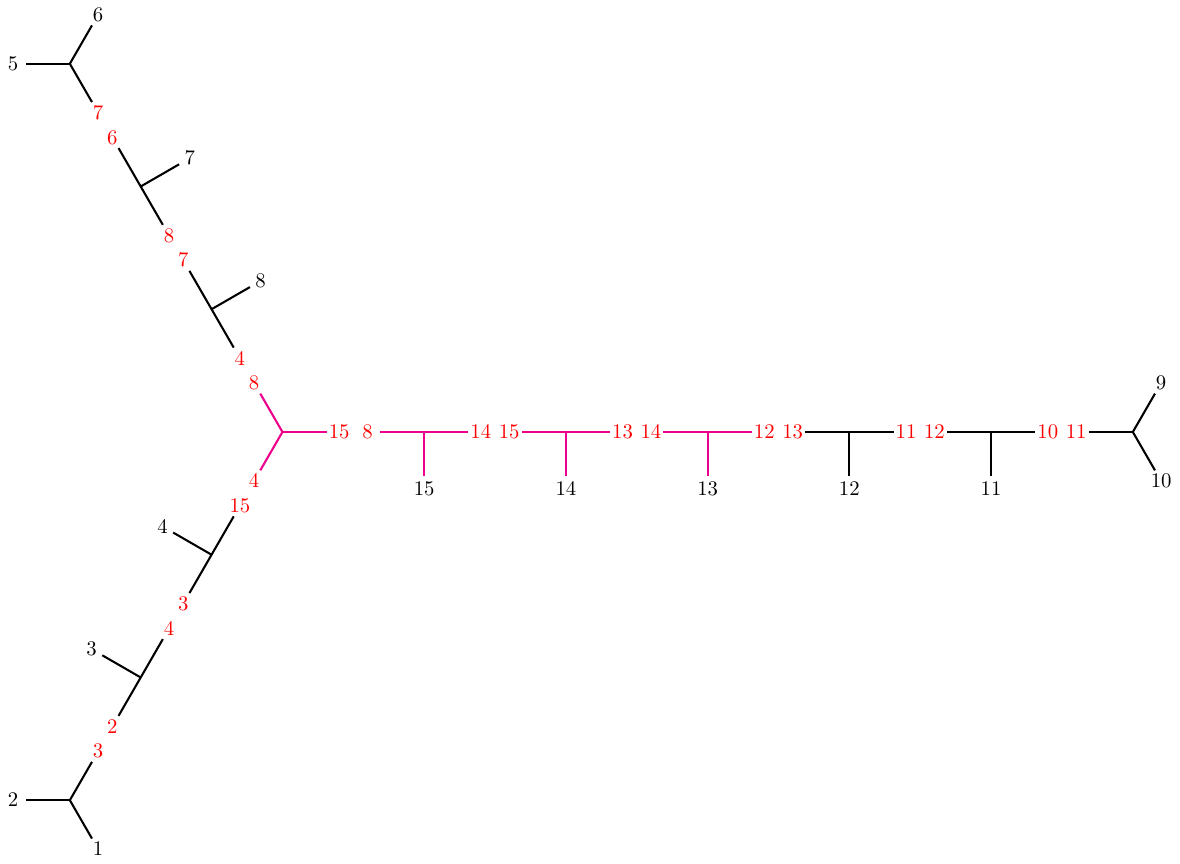}
    \caption{Flow diagram obtained applying to the OPE diagram in Figure~\ref{fig:15-points} the labeling rules explained in this section. The highlighted part of the diagram corresponds to a six-point substructure of the type we will discuss in Section~\ref{sect:six-pt-substructures}. The Casimir operator associated with the leg at the center of this substructure (the one between 14 and 15), can both be written as $\mathcal{D}_{1\dots8,15}^{(\chan_{15})}$ or as $\mathcal{D}_{9\dots14}^{(\chan_{15})}$ and it will be referred to in the section mentioned above.}
    \label{fig:15-point-flow}
\end{figure}

To produce one such type of consistent labeling for an OPE channel $\chan$, one can start by drawing its decomposition into three-point vertices writing explicitly only the labels for the external legs (\textit{e.g.}\ only the black labels in Figures~\ref{fig:five-pt-flow} and~\ref{fig:15-point-flow}), and then fill the labels for internal legs using the three following simple rules:
\begin{itemize}
    \item For any vertex that contains two external leg labels, write the highest\footnote{For sake of symmetry, we contravene to this rule on one of the extremes of the comb-channel case. In any case, a change of this sort corresponds to a $\mathbb{Z}_2$ flip of the OPE diagram in question, and thus does not lead to any substantial difference in the final results.} of these two numbers in the only internal leg that is pointing to this vertex, \textit{e.g.}\
    
    \begin{center}\begin{tikzpicture}[baseline={([yshift=-0.6 ex]current bounding box.center)},scale=2]
        \draw (0,0) -- (0.5,0);
        \draw (0.25,0) -- (0.25,0.25);
        \draw (1.1,0) -- (1.6,0);
        \draw (1.35,0) -- (1.35,0.25);
            \draw[anchor=east] node at (0,0) {\framebox{$1$}};
            \draw[anchor=south] node at (0.25,0.25) {\framebox{$2$}};
            \draw[anchor=west] node at (0.5,0) {$.$};
            \draw[anchor=east] node at (1.1,0) {\colorbox{back-clr}{\color{red}$2$}};
            \draw[anchor=south] node at (1.35,0.25) {$.$};
            \draw[anchor=west] node at (1.6,0) {$.$};
    \end{tikzpicture}
    \end{center}
    \item For any vertex that contains one external leg label, propagate that label to both internal legs that are connected to this vertex, \textit{e.g.}\ 

    \begin{center}\begin{tikzpicture}[baseline={([yshift=-0.6 ex]current bounding box.center)},scale=2]
        \draw (0,0) -- (0.5,0);
        \draw (0.25,0) -- (0.25,0.25);
        \draw (1.1,0) -- (1.6,0);
        \draw (1.35,0) -- (1.35,0.25);
        \draw (2.2,0) -- (2.7,0);
        \draw (2.45,0) -- (2.45,0.25);
            \draw[anchor=east] node at (0,0) {$.$};
            \draw[anchor=south] node at (0.25,0.25) {$.$};
            \draw[anchor=west] node at (0.5,0) {\colorbox{back-clr}{{\color{red}$3$}}};
            \draw[anchor=east] node at (1.1,0) {$.$};
            \draw[anchor=south] node at (1.35,0.25) {\framebox{$3$}};
            \draw[anchor=west] node at (1.6,0) {$.$};
            \draw[anchor=east] node at (2.2,0) {\colorbox{back-clr}{{\color{red}$3$}}};
            \draw[anchor=south] node at (2.45,0.25) {$.$};
            \draw[anchor=west] node at (2.7,0) {$.$};
    \end{tikzpicture}
    \end{center}
    \item For any vertex that has no external leg labels, any pair of internal leg labels propagates the furthest label in clockwise order to the neighboring vertex that is not connected to any of those two internal legs, \textit{e.g.}\ 
    \begin{center}\begin{tikzpicture}[baseline={([yshift=-0.6 ex]current bounding box.center)},scale=2]
        \draw (-0.05,0) -- (0.45,0);
        \draw (0.2,0) -- (0.2,0.25);
        \draw (1.1,0) -- (1.6,0);
        \draw (1.35,0) -- (1.35,0.25);
        \draw (2.25,0) -- (2.75,0);
        \draw (2.5,0) -- (2.5,0.25);
        \draw (1.35,0.95) -- (1.35,1.2);
        \draw (1.13,1.3) -- (1.35,1.2);
        \draw (1.57,1.3) -- (1.35,1.2);
            \draw[anchor=east] node at (-0.05,0) {$.$};
            \draw[anchor=south] node at (0.2,0.25) {$.$};
            \draw[anchor=west] node at (0.45,0) {{\colorbox{back-clr}{\color{red}$6$}}};
            \draw[anchor=east] node at (1.1,0) {\framebox{\color{red}$4$}};
            \draw[anchor=south] node at (1.35,0.25) {\framebox{\color{red}$5$}};
            \draw[anchor=west] node at (1.6,0) {\framebox{\color{red}$6$}};
            \draw[anchor=east] node at (2.25,0) {{\colorbox{back-clr}{\color{red}$5$}}};
            \draw[anchor=south] node at (2.5,0.25) {$.$};
            \draw[anchor=west] node at (2.75,0) {$.$};
            \draw[anchor= west] node at (1.57,1.33) {$.$};
            \draw[anchor= east] node at (1.13,1.33) {$.$};
            \draw[anchor= north] node at (1.35,0.95) {{\colorbox{back-clr}{\color{red}$4$}}};
    \end{tikzpicture}
    \end{center}
\end{itemize}
Applying this procedure to the 15-point function of Figure~\ref{fig:15-points} gives the flow diagram of Figure~\ref{fig:15-point-flow}.

Once a consistent labeling of this type is done, we can easily identify lower-point structures within the higher-point channel, and use this to our advantage to construct conventions for cross-ratios and conformal prefactors that are as compatible as possible with OPE limits.

Given that in $d=1$ the number of cross-ratios is equal to the number of internal legs of a diagram, and that every internal leg is in 1:1 correspondence with a pair of neighboring vertices that make up a four-point function, we can then conventionally choose to work with a set of four-point cross-ratios constructed out of every pair of neighboring vertices in the following way:
\begin{equation}
    z_{(abcd)}\equiv z\left({\vspace*{-100pt}\begin{tikzpicture}[baseline={([yshift=-0.6 ex]current bounding box.center)}]
        \draw (0,0) -- (0.5,0);
        \draw (0.25,0) -- (0.25,0.25);
        \draw (1.2,0) -- (1.7,0);
        \draw (1.45,0) -- (1.45,0.25);
        \begin{scriptsize}
            \draw[anchor=east] node at (0,0) {$a$};
            \draw[anchor=south] node at (0.25,0.25) {$b$};
            \draw[anchor=west] node at (0.5,0) {$c$};
            \draw[anchor=east] node at (1.2,0) {$b$};
            \draw[anchor=south] node at (1.45,0.25) {$c$};
            \draw[anchor=west] node at (1.7,0) {$d$};
        \end{scriptsize}
    \end{tikzpicture} }\right)= \frac{x_{ab}\,x_{cd}}{x_{ac}\,x_{bd}}\,,
    \label{eq:general_cross_ratio}
\end{equation}
where the legs $a$ and $d$ are identified as those that only appear once in those two vertices.
For the conformal prefactor, we will construct it from the three-point vertices where the external legs appear. For every external leg $i$ we introduce an object
\begin{equation}
    V_i=V\left({\vspace*{-100pt}\begin{tikzpicture}[baseline={([yshift=-0.6 ex]current bounding box.center)}]
        \draw (0,0) -- (0.5,0);
        \draw (0.25,0) -- (0.25,0.25);
        \begin{scriptsize}
            \draw[anchor=east] node at (0,0) {$a$};
            \draw[anchor=south] node at (0.25,0.25) {$i$};
            \draw[anchor=west] node at (0.5,0) {$b$};
        \end{scriptsize}
    \end{tikzpicture} }\right)= \left(\frac{\abs{x_{ab}}}{\abs{x_{a i}}\abs{x_{bi}}}\right)^{h_i} 
\end{equation}
which picks up a ratio of distances between the position $x_i$ and the coordinates $x_a$ and $x_b$ that appear in the three-point vertex where the external leg $i$ is inserted.  With this being defined, our convention for the prefactor will simply be the product of all the $V_i$'s:
\begin{equation}
    \Omega_N^{(f)}= \prod_{i=1}^{N} V_i\,.
\end{equation}

\subsection{Six-point-constructible diagrams and \texorpdfstring{AdS$_3^{\otimes m}$}{AdS3×m} wavefunctions}
\label{sec:six-point-constructible}
Given our conventions stated above, we can now start to construct Casimir operators for channels with various topologies and arbitrary numbers of points. The starting key observation is that using our conventions above, the Casimir operators for comb-channel blocks will always have the same form as those for six-point comb-channel blocks. This observation will then be extended in Section~\ref{sect:six-pt-substructures} to apply also to certain Casimir operators in other topologies. Finally, we use these results to describe a class of conformal blocks that can be directly seen to correspond to AdS$_3^{\otimes m}$ free-particle wavefunctions.

\subsubsection{The Casimir operators for comb-channel blocks}
\label{subsect:Casimirs_comb_general}

\begin{figure}
    \centering
    \includegraphics{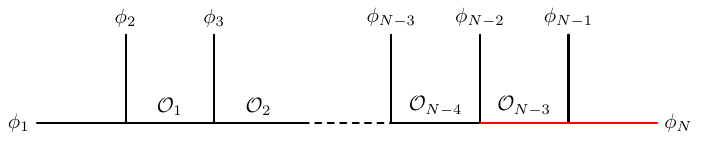}
    \caption{Depiction of a general comb-channel OPE diagram. The legs highlighted in red are those that are introduced when increasing from $N-1$ to $N$ points. In the Casimir operators, only the dependence on the coordinates of fields to the right of the dashed line is affected by this increase.}
    \label{fig:general_comb}
\end{figure}
With our conventions for prefactors and cross-ratios, we have the following recursive definition
\begin{equation}
    \Omega_N^{\{h_i\}}=\Omega_{N-1}^{\{h_i\}} z_{N-3}^{h_{N-1}} \left(\frac{x_{N-2,N-1}}{x_{N-2,N}x_{N-1,N}}\right)^{h_N}\,.
    \label{eq:pref_N_as_N-1}
\end{equation}
When going from $N-1$ to $N$ points, we can observe that the dependence on the coordinates $x_1,\dots,x_{N-4}$ is left untouched, since these only appear in $\Omega_{N-1}^{\{h_i\}}$ in~\eqref{eq:pref_N_as_N-1} and are not present in the new cross-ratio $z_{N-3}$, see also Figure~\ref{fig:general_comb}.
This means that the expressions for the Casimir operators $\widetilde{\mathcal{D}}_{1\dots i}^{(N)}$ with $i\le N-4$ will also be unaltered, since these Casimirs only involve generators acting on points up to $x_{N-4}$. At this point, we can use the knowledge of the six-point Casimir operators and the $\mathbb{Z}_2$-reflection symmetry of the $N$-point comb-channel blocks under the transformation
\begin{equation}
    \mathcal{P}_N:\begin{array}{c} z_i\rightarrow z_{N-2-i} \\
    h_i\rightarrow h_{N+1-i} \end{array}, \qquad i=\{1,2,\dots,N\}\,,
\end{equation}
to write the last two Casimir operators as
\begin{equation}
    \widetilde{\mathcal{D}}_{N-1,N}^{(N)}=\mathcal{P}_N\,\widetilde{\mathcal{D}}_{12}^{(N)}\,, \qquad \widetilde{\mathcal{D}}_{1,\dots,N-3}^{(N)}=\mathcal{P}_N\,\widetilde{\mathcal{D}}_{123}^{(N)}\,.
\end{equation}
But $\widetilde{\mathcal{D}}_{12}^{(N)}=\widetilde{\mathcal{D}}_{12}^{(5)}$ and $\widetilde{\mathcal{D}}_{123}^{(N)}=\widetilde{\mathcal{D}}_{123}^{(6)}$, so the Casimirs at the extremes of the comb will always have the five-point form
\begin{gather}
    \widetilde{\mathcal{D}}_{12}^{(N)}=z_1\left[(1-z_1)z_1\partial_{z_1}^2-z_1 z_2 \partial_{z_1}\partial_{z_2}-(1-a-h_3) z_1 \partial_{z_1} + a z_2 \partial_{z_2} -a h_3\right] \label{eq:GenCas12}\\
    \begin{split}\widetilde{\mathcal{D}}_{(N-1)N}^{(N)}=z_{N-3}\Bigl[(1-z_{N-3})z_{N-3}\partial_{z_{N-3}}^2\!-z_{N-4} z_{N-3} \partial_{z_{N-4}}\partial_{z_{N-3}}\!-(1&-\tilde{a}-h_{N-2}) z_{N-3} \partial_{z_{N-3}} \\
    &+ \tilde{a} z_{N-4} \partial_{z_{N-4}} -\tilde{a} h_{N-2}\Bigr] 
    \end{split}\label{eq:GenCaslast}
\end{gather}
while the more internal Casimirs in the comb can all be seen as transformations of $\widetilde{\mathcal{D}}_{123}^{(6)}$:
\begin{multline}
    \widetilde{\mathcal{D}}_{1\dots i}^{(N)}\!=\!\mathcal{P}_{i+3}\widetilde{\mathcal{D}}_{123}^{(6)}\!=\!z_{i-1}\Bigl[(1-z_{i-1})z_{i-1}\partial_{z_{i-1}}^2\!-z_{i-2} z_{i-1} \partial_{z_{i-2}}\partial_{z_{i-1}}\!-z_{i-2} z_{i} \partial_{z_{i-2}}\partial_{z_{i}}\!-z_{i-1} z_{i} \partial_{z_{i-1}}\partial_{z_{i}}\\
    -(1-h_{i}-h_{i+1}) z_{i-1} \partial_{z_{i-1}} + h_{i} z_{i} \partial_{z_{i}}+ h_{i+1} z_{i-2} \partial_{z_{i-2}} -h_{i} h_{i+1}\Bigr],
    \label{eq:GenCasmiddle}
\end{multline}
since with this recursive construction the operator $\widetilde{\mathcal{D}}_{1\dots i}^{(N)}=\widetilde{\mathcal{D}}_{1\dots i}^{(i+3)}$ was produced by $\mathbb{Z}_2$-reflection of $\widetilde{\mathcal{D}}_{123}^{(i+3)}=\widetilde{\mathcal{D}}_{123}^{(6)}$ when considering an $(i+3)$-point function. For some explicit checks of these expressions, we invite the reader to use the \texttt{Mathematica} notebook provided as supplemental material of this publication.
This completes the derivation of all Casimir operators relevant for one- and two-dimensional comb-channel blocks.

\subsubsection{Six-point sub-structures}
\label{sect:six-pt-substructures}
\begin{figure}[htp]
    \centering
    \includegraphics{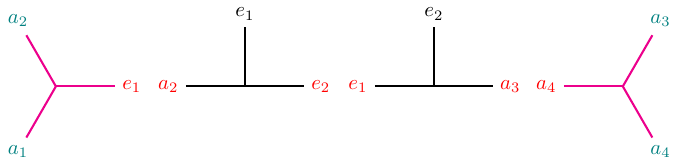}
    \caption{Six-point substructure that can appear within a flow diagram. The $e_i$ labels correspond to external legs, while the $a_j$ labels can correspond to either internal or external legs. Whenever a structure like that of this figure is present in a flow diagram, the Casimir operator $\widetilde{\mathcal{D}}_{p\dots e_1}^{(\chan_N)}$ associated with the internal leg located between $e_1$ and $e_2$ can be related to the AdS$_3$ free Hamiltonian.}
    \label{fig:six-point-substructure}
\end{figure}

Let us now turn to another crucial observation: individual Casimirs do not carry information about the topology of the OPE diagram in which they are located. To have information that fully specifies the topology of the OPE diagram one is considering, one needs in fact the whole set of Casimir operators. This means that, with appropriate conventions for prefactors and cross-ratios, any Casimir operator $\widetilde{\mathcal{D}}_{p\dots q}^{(\chan_N)}$ can acquire the same form of the analogous Casimir $\widetilde{\mathcal{D}}_{p\dots q}^{(N_p)}$ for a corresponding comb-channel topology. We will now argue that our conventions are such that, given an OPE channel $\chan_N$ with flow diagram $f_{\chan_N}$, every time one encounters in $f_{\chan_N}$ a sub-structure that looks like a six-point comb as in Figure~\ref{fig:six-point-substructure}
\!\!---where $e_1$ and $e_2$ are external legs while the $a_i$ can be either internal or external---then the Casimir operator $\widetilde{\mathcal{D}}_{p\dots e_1}^{(\chan_N)}$ associated with the internal leg located between $e_1$ and $e_2$ is in our conventions automatically of the form of the middle-leg six-point Casimir $\widetilde{\mathcal{D}}_{123}^{(6)}$, \textit{i.e.}\ the AdS$_3$ free Hamiltonian once we extract an appropriate prefactor. 

In fact, from what we mentioned above, we know already that this will be true \emph{in some coordinates}, as one can embed the same $\widetilde{\mathcal{D}}_{p\dots e_1}^{(\chan_N)}$ within a comb-channel diagram $(N')$ with $p,\dots, e_1$ all located on the same side of the comb, without any additional legs in between. In short,
\begin{equation}
    \widetilde{\mathcal{D}}_{p\dots e_1}^{(\chan_N)}=\widetilde{\mathcal{D}}_{p\dots e_1}^{(N')} \quad \text{in some coordinates.}
\end{equation} 
Now we can imagine performing $N-6$ OPE limits from our initial channel $\chan_N$ to reduce to just the six-point channel $(6')$ whose flow diagram $f_{(6')}$ is directly what is represented in Figure~\ref{fig:six-point-substructure}, with $a_i$ all external legs. The middle-leg Casimir $\widetilde{\mathcal{D}}_{a_1a_2e_1}^{(6')}$ for this OPE-reduced block will be identical to the $\widetilde{\mathcal{D}}_{p\dots e_1}^{(N')}$ Casimir of the $(N')$ comb channel, given what we showed in Section~\ref{subsect:Casimirs_comb_general}. We can now say that since 
\begin{enumerate}
    \item $\widetilde{\mathcal{D}}_{p\dots e_1}^{(\chan_N)}$ has the form of a comb-channel middle-leg Casimir $\widetilde{\mathcal{D}}_{p\dots e_1}^{(N')}$ in some coordinates,
    \item the OPE limit of $\widetilde{\mathcal{D}}_{p\dots e_1}^{(\chan_N)}$ in our coordinates is directly equal to the middle-leg Casimir of a six-point comb channel $\widetilde{\mathcal{D}}_{a_1a_2e_1}^{(6')}$, itself equal to the higher-point comb Casimir $\widetilde{\mathcal{D}}_{p\dots e_1}^{(N')}$,
    \item the three cross-ratios of the OPE-reduced block $\tilde{\psi}^{(6')}$ are also present in the initial set of cross ratios of the channel $\chan_N$ due to our conventions for~\eqref{eq:general_cross_ratio},
\end{enumerate}
we conclude that in our choice of coordinates $\widetilde{\mathcal{D}}_{p\dots e_1}^{(\chan_N)}= \widetilde{\mathcal{D}}_{p\dots e_1}^{(N')}$ also outside of the OPE limit. For instance, the Casimir $\widetilde{\mathcal{D}}_{9\dots 14}^{(\chan_{15})}$ of the 15-point function in Figure~\ref{fig:15-point-flow} in our conventions will acquire precisely the same form of the $\widetilde{\mathcal{D}}_{9\dots 14}^{(15')}$ Casimir of a comb channel that shares the same six-point substructure highlighted in magenta in that figure.
Explicitly, $\widetilde{\mathcal{D}}_{9\dots 14}^{(\chan_{15})} = \widetilde{\mathcal{D}}_{9\dots 14}^{(15')} = \widetilde{\mathcal{D}}_{12,13,14}^{(6')}$ for the six-point substructure in magenta in Figure~\ref{fig:15-point-flow}.

In order to make manifest the AdS$_3$ Hamiltonian, we need to extract the generalized analog of the conjugating prefactor $\omega_6^{\{h_i\}}$ presented in~\eqref{eq:Ham6Comb_and_pref}. For the six-point sub-structure of Figure~\ref{fig:six-point-substructure} it will simply be
\begin{equation}
    \omega_{(a_1a_2e_1e_2a_3a_4)}^{\{h_i\}}=z_{(a_1a_2e_1e_2)}^{h_{e_1}}z_{(e_1e_2a_3a_4)}^{h_{e_2}}\,.
    \label{eq:conjugation_prefactor_substructure}
\end{equation}
The observation made in this subsection is much deeper than what it may initially seem: it gives us a recipe for computing conformal blocks of a certain class we will now discuss.

\subsubsection{Gluing six-point building blocks}
\label{subsect:Gluing_six-pt}

Let us consider a $N=(3+3m)$-point diagram with $m>0$.  We will define it to be \emph{six-point constructible} (6PC) if it can be obtained by taking $m$ copies of six-point comb-channel diagrams as the one in Figure~\ref{fig:six-point-substructure}, and ``gluing'' them together via identification of the vertices on their extremes (highlighted in magenta in Figure~\ref{fig:six-point-substructure}). 
From the point of view of differential operators, this gluing procedure can be thought of as considering a function space that is acted upon by $m$ collections of generators (labelled with Latin indices {\it e.g.} $i,j,a,b$) associated with six-point configurations $\{\mathcal{T}_{(i_n)}^{\alpha}\}$---with conformal Ward identities $\sum_{n=1}^6 \mathcal{T}_{(i_n)}^{\alpha} =0$ satisfied for each individual collection when acting on this function space---and then identifying within this function space the action of certain (sums of) generators of one collection, with (sums of) generators of another $\sum_{n\in S_i} \mathcal{T}_{(i_n)}^{\alpha} =\sum_{\ell \in S_j} \mathcal{T}_{(j_\ell)}^{\alpha}$, where $S_i, S_j$ are subsets of $\{1,2,\ldots,6\}$. The conformal blocks $\tilde{\psi}^{(\chan_N)}$ will be particular functions within this function space.

\smallskip

\noindent Concretely, there are two types of gluing one can perform:
\begin{itemize}
    \item Gluing of a pair of diagrams
    \begin{center}
        \includegraphics[scale=0.75]{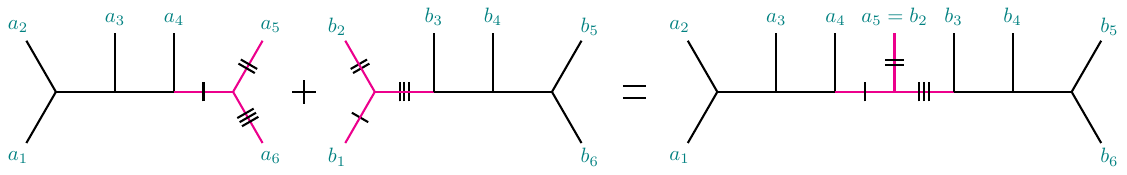}
    \end{center}
    which at the level of the generators can be written as
    \begin{equation}
        \mathcal{T}^{\alpha}_{(b_1)}\equiv \mathcal{T}^{\alpha}_{(a_5)}+\mathcal{T}^{\alpha}_{(a_6)}\,, \qquad \mathcal{T}^{\alpha}_{(a_5)}\equiv \mathcal{T}^{\alpha}_{(b_2)}\,, \qquad \mathcal{T}^{\alpha}_{(a_6)} \equiv \mathcal{T}^{\alpha}_{(b_1)}+\mathcal{T}^{\alpha}_{(b_2)}\,.
    \end{equation}
    \item Gluing of a triplet of diagrams 
    \begin{center}
        \includegraphics[scale=0.74]{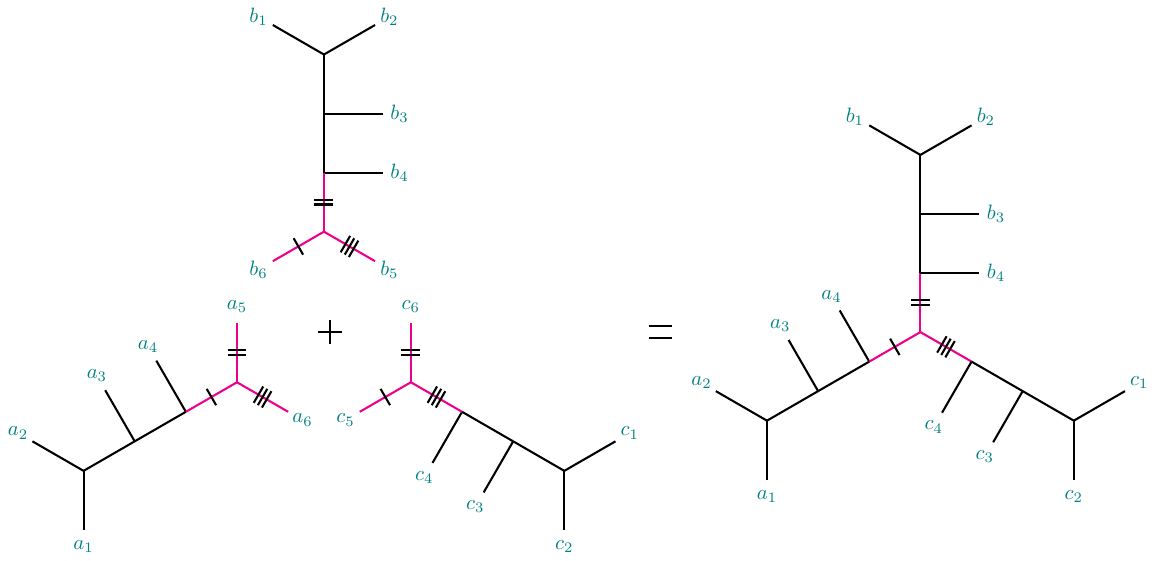}
    \end{center}
    which at the level of generators can be written as
    \begin{equation}
            \mathcal{T}^{
            \alpha}_{(a_5)}+\mathcal{T}^{
            \alpha}_{(a_6)}\!\equiv \mathcal{T}^{
            \alpha}_{(b_6)}\!\equiv \mathcal{T}^{
            \alpha}_{(c_5)},\quad  \mathcal{T}^{
            \alpha}_{(a_5)}\!\equiv \mathcal{T}^{
            \alpha}_{(b_5)}+\mathcal{T}^{
            \alpha}_{(b_6)}\!\equiv \mathcal{T}^{
            \alpha}_{(c_6)}, \quad \mathcal{T}^{
            \alpha}_{(a_6)}\!\equiv \mathcal{T}^{
            \alpha}_{(b_5)}\!\equiv \mathcal{T}^{
            \alpha}_{(c_5)}+\mathcal{T}^{
            \alpha}_{(c_6)}.
    \end{equation}
    \end{itemize}
    In both cases, the conformal Ward identities of the individual six-point functions construct the conformal Ward identities of the higher-point function:
    \begin{equation}
        \mathcal{T}^{\alpha}_{1}+\mathcal{T}^{\alpha}_{2}+\dots \mathcal{T}^{
            \alpha}_{N}=0\,.
    \end{equation}
If we restrict our attention to diagrams that are 6PC, we then know from the start one basis of functions that span the space where the conformal blocks lie. From what we observed in Section~\ref{sect:six-pt-substructures}, we have in fact that every six-point sub-structure describes a copy of AdS$_3$ and includes its free Hamiltonian as the middle-leg Casimir once one extracts an appropriate prefactor. For any generic 6PC diagram, built out of $m$ six-point sub-structures labeled by $s=1,\dots,m$, the prefactors $\omega_{[s]}^{\{h_i\}}$ needed to be extracted for each individual six-point building block are all compatible with each other, as they all involve different cross-ratios, cf.~\eqref{eq:conjugation_prefactor_substructure} and~\eqref{eq:general_cross_ratio}. This means that we can construct the free AdS$_3$ Hamiltonian for every six-point building block, and thus 6PC diagrams give rise to a system that lives in a tensor product of $(N-3)/3$ AdS$_3$ spaces. The full Hamiltonian of the system is simply the sum of all the AdS$_3$ Hamiltonians $\mathcal{H}^{[s]}$ with $s$ labeling the factors of the tensor product:
\begin{equation}
    \mathcal{H}_{\mathrm{tot}}=\sum_{s=1}^{\frac{N-3}{3}} \mathcal{H}^{[s]}\,.
\end{equation}
There is actually no concrete advantage in considering this operator $\mathcal{H}_{\mathrm{tot}}$ over the individual $\mathcal{H}^{[s]}$'s, so we will keep on referring to this system as being made of many AdS$_3$ free Hamiltonians. 
What we just derived means that we can write down conformal blocks for 6PC diagrams as linear combinations of wave modes in AdS$_3^{\otimes \frac{N-3}{3}}$, where the precise coefficients can be found by acting with the Casimirs that do not sit at the middle legs of the six-point building blocks, \textit{i.e.}\ the commuting charges $Q_i^{[s]}$.

As we saw in  Section~\ref{sect:plane_waves_CB}, the action of the two conserved charges $Q_1$ and $Q_3$ associated with a certain AdS$_3$ leads to two decoupled recurrence relations, as the two operators can be constructed out of two different sets of generators $J^1$ and $J^3$ that commute one with the other. In other words, the Hamiltonian divides the diagram into two parts, and the solution of each part is decoupled from the solution of the other. Since the same structure is present in the 6PC diagrams, this claim extends also to these, \textit{i.e.}\ diagonalizations of operators that are separated by AdS$_3$ Hamiltonians within an OPE channel can be performed independently. With this in mind, the solution to determining general 6PC conformal blocks becomes simply the study of which structures can be enclosed between two AdS$_3$ Hamiltonians, or between one AdS$_3$ Hamiltonian and an edge of the OPE diagram. 
We have just three types of structures that can be produced:
\begin{enumerate}
    \item There is nothing glued to one side of the six-point building block, which has to then sit at one edge of the diagram, \textit{e.g.}\
    \begin{center}
        \includegraphics[scale=0.75]{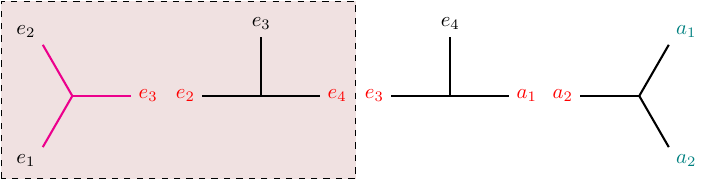}
    \end{center}
    The simplest example in which this happens is directly the six-point comb-channel blocks, which obviously just contain one six-point sub-structure that is not glued to anything on either side. Every time we encounter one such edge of the diagram, there is only one Casimir operator $\widetilde{\mathcal{D}}_{e_1e_2}^{(\chan_N)}$ to diagonalize, which has the usual comb-channel form~\eqref{eq:D12_sixpt_comb}, and thus its diagonalization works exactly the same as the one for the six-point blocks we performed in Section~\ref{sect:plane_waves_CB}. Focusing on the two cross-ratios $z_{(e_1e_2e_3e_4)}$ and $z_{(e_2e_3e_4a_1)}$ that the Casimir operator depends on---and renaming them for brevity as $z_{b_1}$ and $z_{b_2}$, respectively, while renaming the other cross ratios $z_{b_r}$ with $r = 3,4,\dots, (N-3)$---we have that $\mathcal{D}_{e_1e_2}^{(\chan_N)}$ is diagonalized by the combination of wave modes
    \begin{equation}
        \psi^{(\chan_N)}\!(z_i)=\sum_{n_{b_1}}\alpha_{n_{b_1}}\! \varphi_{h_{b_1}+n_{b_1}-h_{e_3},h_{b_2},h_{b_3}+n_{b_3}-h_{e_4}}(z_{b_1},z_{b_2},z_{b_3})
         f(z_{b_3},\dots,z_{b_{N-3}})
        \label{eq:edge_diagram_block}
    \end{equation}
    with $f$ arbitrary function of the remaining cross-ratios, and $\alpha_{n_{b_1}}$ corresponding to the combination coefficients
    \begin{equation}
        \alpha_{n_{b_1}}=\alpha^{(\mathbbm{h}_{b_1},\mathbbm{h}_{b_2},h_{e_1},h_{e_2},h_{e_3})}_{n_{b_1}}=\frac{\left(\mathbbm{h}_{b_1}+\mathbbm{h}_{b_2}-h_{e_3}\right)_{n_{b_1}}\left(\mathbbm{h}_{b_1}-h_{e_2}+h_{e_1} \right)_{n_{b_1}}}{n_{b_1}!\left(2 \mathbbm{h}_{b_1}\right)_{n_{b_1}}}\,,
    \end{equation}
    which are nothing but~\eqref{eq:coefficients_six_points} adapted to a general diagram.
    \item There can be a gluing of two six-point sub-structures in a comb-like fashion, thus having a gluing vertex involving one external leg and two internal ones
    \begin{center}
        \includegraphics[scale=0.75]{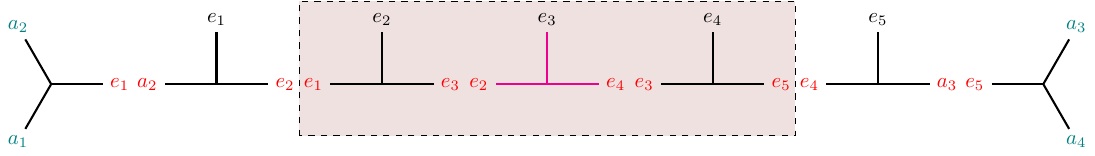}
    \end{center}
    The simplest example in which this happens is in a nine-point comb-channel block, which is made of two six-point building blocks. To perform the gluing one needs to diagonalize the two Casimir operators for the internal legs between $e_2$ and $e_4$, which when acting on a linear combination of wave modes~\eqref{eq:AdS_Plane_waves}
    \begin{multline}
    \varphi_{n_{b_1},n_{b_3}}\varphi_{n_{b_4},n_{b_6}} \equiv\\
    \varphi_{\mathbbm{h}_{b_1}-h_{e_1}+n_{b_1},\mathbbm{h}_{b_2},\mathbbm{h}_{b_3}-h_{e_2}+n_{b_3}}(z_{b_1},z_{b_2},z_{b_3}) \varphi_{\mathbbm{h}_{b_4}-h_{e_4}+n_{b_4},\mathbbm{h}_{b_5},\mathbbm{h}_{b_6}-h_{e_5}+n_{b_6}}(z_{b_4},z_{b_5},z_{b_6})
    \end{multline} 
    of the form
    \begin{equation}
        \sum_{\{n_{b_i}\!\}} \beta_{n_{b_3},n_{b_4}}\varphi_{n_{b_1},n_{b_3}}\varphi_{n_{b_4},n_{b_6}} f(z_{b_7},\dots z_{b_{N-3}})
    \end{equation}
    provide the following recurrence relation for $\beta_{n_{b_3},n_{b_4}}$
    \begin{equation}
        \beta _{n_{b_3}+1,n_{b_4}}=\frac{\left(\mathbbm{h}_{b_2}+\mathbbm{h}_{b_3}-h_{e_2}+n_{b_3}\right) \left(\mathbbm{h}_{b_3}+\mathbbm{h}_{b_4}-h_{e_3}+n_{b_3}+n_{b_4}\right)}{(n_{b_3}+1) \left(2 \mathbbm{h}_{b_3}+n_{b_3}\right) }\beta _{n_{b_3},n_{b_4}}
    \end{equation}
as well as a $\mathbb{Z}_2$-reflected one for the shift in $n_{b_4}$. The solution of the two recurrence relations is given by
\begin{equation}
    \beta_{n_{b_3},n_{b_4}}=\frac{\left(\mathbbm{h}_{b_2}+\mathbbm{h}_{b_3}-h_{e_2}\right)_{n_{b_3}} \left(\mathbbm{h}_{b_3}+\mathbbm{h}_{b_4}-h_{e_3}\right)_{n_{b_3}+n_{b_4}} \left(\mathbbm{h}_{b_4}+\mathbbm{h}_{b_5}-h_{e_4}\right)_{n_{b_4}}}{n_{b_3}! n_{b_4}! \left(2 \mathbbm{h}_{b_3}\right)_{n_{b_3}} \left(2 \mathbbm{h}_{b_4}\right)_{n_{b_4}}}\,.
    \end{equation}
    \item There can be a gluing of three six-point building blocks with a vertex that involves only internal legs. 
    \begin{center}
        \includegraphics[scale=0.7]{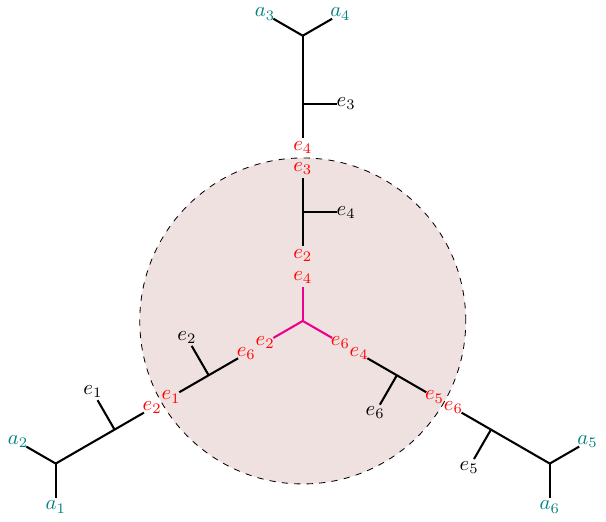}
    \end{center}
    The simplest example in which this happens is a twelve-point function in an ``extended snowflake'' channel. In this case, there are three Casimir equations to be solved, whose diagonalization when acting on the product of wave modes 
    \begin{multline}
    \varphi_{n_{b_1},n_{b_3}}\varphi_{n_{b_4},n_{b_6}}\varphi_{n_{b_7},n_{b_9}} \equiv\\
    \varphi_{\mathbbm{h}_{b_1}-h_{e_1}+n_{b_1},\mathbbm{h}_{b_2},\mathbbm{h}_{b_3}-h_{e_2}+n_{b_3}}(z_{b_1},z_{b_2},z_{b_3}) \varphi_{\mathbbm{h}_{b_4}-h_{e_3}+n_{b_4},\mathbbm{h}_{b_5},\mathbbm{h}_{b_6}-h_{e_4}+n_{b_6}}(z_{b_4},z_{b_5},z_{b_6})\\
    \times \varphi_{\mathbbm{h}_{b_7}-h_{e_5}+n_{b_7},\mathbbm{h}_{b_8},\mathbbm{h}_{b_9}-h_{e_6}+n_{b_9}}(z_{b_7},z_{b_8},z_{b_9}) 
    \end{multline} 
    with $z_{b_3},z_{b_6},z_{b_9}$ the cross-ratios constructed around the central vertex, amounts to computing the coefficients    $\gamma_{n_{b_3},n_{b_6},n_{b_9}}$ of the combination
    \begin{equation}
        \sum_{\{n_{b_i}\!\}} \gamma_{n_{b_3},n_{b_6},n_{b_9}}\varphi_{n_{b_1},n_{b_3}}\varphi_{n_{b_4},n_{b_6}}\varphi_{n_{b_7},n_{b_9}} f(z_{b_{10}},\dots z_{b_{N-3}}),
    \end{equation}
    where $f$ is again an arbitrary function of its arguments.
    In our set of conventions, the problem is manifestly cyclically symmetric, with recurrence relations
    \begin{multline}
        \left(\mathbbm{h}_{b_2}+\mathbbm{h}_{b_3}-h_{e_2}+n_{b_3}\right) \left(\mathbbm{h}_{b_5}+\mathbbm{h}_{b_6}-h_{e_4}+n_{b_6}\right)\gamma _{n_{b_3},n_{b_6},n_{b_9}} \\
        -\left(\mathbbm{h}_{b_3}+\mathbbm{h}_{b_6}-\mathbbm{h}_{b_9}+n_{b_3}+n_{b_6}-n_{b_9}+1\right) \left(\mathbbm{h}_{b_2}+\mathbbm{h}_{b_3}-h_{e_2}+n_{b_3}\right) \gamma _{n_{b_3},n_{b_6}+1,n_{b_9}} \\
        +\left(n_{b_3}+1\right) \left(2 \mathbbm{h}_{b_3}+n_{b_3}\right) \gamma _{n_{b_3}+1,n_{b_6}+1,n_{b_9}}=0
    \end{multline}
    and its cyclically transformed versions.
    The solution to this set of recurrence relations is
    \begin{multline}
        \gamma_{n_{b_3},n_{b_6},n_{b_9}}\!=\frac{1}{n_{b_3}!\,n_{b_6}!\,n_{b_9}!}\frac{\left(\mathbbm{h}_{b_2}\!+\mathbbm{h}_{b_3}\!-h_{e_2}\right)_{n_{b_3}}}{\left(2 \mathbbm{h}_{b_3}\right)_{n_{b_3}}}\frac{\left(\mathbbm{h}_{b_5}\!+\mathbbm{h}_{b_6}\!-h_{e_4}\right)_{n_{b_6}}}{\left(2 \mathbbm{h}_{b_6}\right)_{n_{b_6}}}\frac{\left(\mathbbm{h}_{b_8}\!+\mathbbm{h}_{b_9}\!-h_{e_6}\right)_{n_{b_9}}}{\left(2 \mathbbm{h}_{b_9}\right)_{n_{b_9}}} \\ 
        \times \left(\mathbbm{h}_{b_3b_6;b_9}-n_{b_9}\right)_{n_{b_3}}\left(\mathbbm{h}_{b_6b_9;b_3}-n_{b_3}\right)_{n_{b_6}}\left(\mathbbm{h}_{b_9b_3;b_6}-n_{b_6}\right)_{n_{b_9}}  \\ 
        \times {}_4F_3\left(\begin{array}{c}
        -n_{b_3}, \, -n_{b_6},\, -n_{b_9},\, \mathbbm{h}_{b_3}+\mathbbm{h}_{b_6}+\mathbbm{h}_{b_9}-1  \\
        \mathbbm{h}_{b_3b_6;b_9}-n_{b_9}\,,\, \mathbbm{h}_{b_6b_9;b_3}-n_{b_3},\mathbbm{h}_{b_9b_3;b_6}-n_{b_6} 
    \end{array}; 1
    \right),
    \label{eq:expr_coeff}
    \end{multline}
    where
    \begin{equation}
    \mathbbm{h}_{ij;k}=\mathbbm{h}_{i}+\mathbbm{h}_{j}-\mathbbm{h}_{k}\,.
\end{equation}
    
\end{enumerate}
With these rules in place, it is now straightforward to compute any 6PC conformal block as linear combinations of AdS$_3^{\otimes \frac{N-3}{3}}$ wave modes. 
The main claim of this paper is now just one tiny step away.

\subsection{Non-6PC diagrams and limits of AdS free wavefunctions}
\label{subsect:non_6PC_diagrams}
To finalize the main claim of this paper, we now need to note that any non-6PC diagram can be embedded in a 6PC one just by adding extra legs to the original diagram. This has the direct implication that we can work out blocks for non-6PC cases as either OPE limits or identity limits of 6PC blocks. We are then ready to state: \emph{$N$-point conformal blocks in one-dimensional CFTs can always be seen as either AdS$_3^{\otimes (\lceil N/3\rceil-1 )}$ free-particle wavefunctions, or limits thereof}. This product space can either be seen as a physical space in which a single free particle is propagating, or as the configuration space of $(\lceil N/3\rceil-1 )$ distinguishable free particles that live in one single copy of AdS$_3$. The conserved charges $Q_i^{[s]}$ that couple different AdS$_3$ spaces in the first interpretation, couple instead two or three particles in the second interpretation. We will take the first point of view when discussing the OPE- or identity-limit reductions, leaving comments on the second interpretation in brackets, where relevant.

To see how the reduction works concretely, we will now analyze both limiting approaches and their concrete applications. The OPE limits are a much simpler tool to use, but cannot be used to derive blocks in all possible cases. The identity limits, instead, are more convoluted in concrete applications, but can always be used to reduce to any diagram starting from a 6PC one.

\subsubsection{OPE limits}
Whenever the diagram associated with the conformal blocks we want to compute differs from a 6PC one only in the length of one or more combs that end with two external legs, the simplest way to reduce 6PC blocks to those for this diagram is to take OPE limits. 

Given a certain comb that needs to be shortened, it only makes sense to consider up to two limits: a third consecutive OPE limit would correspond in fact to just removing one six-point sub-structure from the original diagram, and thus removing one AdS$_3$ from the tensor-product space (or one particle, if interpreting as a multi-particle system).

We then just need to focus on one edge of the diagram, which is associated with two cross-ratios $z_{b_1}$ and $z_{b_2}$ that belong to the $\bar{s}$-th copy of AdS$_3$ in the tensor product ($\bar{s}$-th particle in AdS$_3$), and whose dependence in the conformal blocks we worked out in Section~\ref{subsect:Gluing_six-pt} 
\begin{multline}
        \tilde{\psi}^{(\chan_N)}\!(z_i)=\sum_{n_{b_1}}\alpha_{n_{b_1}}\! z_{b_1}^{\mathbbm{h}_{b_1}+n_{b_1}}\!z_{b_2}^{\mathbbm{h}_{b_2}}{}_2F_1\!\left(\mathbbm{h}_{b_1}\!+\!\mathbbm{h}_{b_2}\!+\!n_{b_1}\!-\!h_{e_3},\mathbbm{h}_{b_2}\!+\!\mathbbm{h}_{b_3}\!+\!n_{b_3}\!-\!h_{e_4},2\mathbbm{h}_{b_2};z_{b_2} \right)
        \\
        \times f(z_{b_3},\dots,z_{b_{N-3}})\,.
\end{multline}
The first OPE limit one can perform here---between the two external fields that sit at the extreme of the comb---corresponds to $z_{b_1}\to 0$. It is straightforward to see how this is implemented by just considering the $n_{b_1}=0$ term within the sum to produce the $z_{b_1}^{\mathbbm{h}_{b_1}}$ asymptotics of the OPE limit. Similarly, a second consecutive OPE limit corresponds to $z_{b_2}\to 0$, which reduces the ${}_2F_1$ to 1, and leads to the $z_{b_2}^{\mathbbm{h}_{b_2}}$ asymptotics. 

From the AdS perspective, the $z_{b_1} \to 0$ limit corresponds to exactly the same type of limit we performed in Section~\ref{ssec:5-pt-Hamiltonian}, and reduces the $\bar{s}$-th AdS$_3$ space to an AdS$_2$ with a background magnetic field (constrains just one of the particles to lie in an AdS$_2$ subspace). The second $z_{b_2}\to 0$ limit, instead, corresponds to approaching the AdS$_2$ boundary at $\rho \to 0$---see also~\eqref{eq:to_conf_flat_five} and~\eqref{eq:five-pt-conf_flat_ops}---which trivializes the magnetic AdS$_2$ Hamiltonian
\begin{equation}
    z_{b_2}^{h_{e_{3}}-\mathbbm{h}_{b_2}}\mathcal{H}^{(\bar{s})}\!\left[z_{b_2}^{\mathbbm{h}_{b_2}-h_{e_{3}}} f(z_{b_3},\dots,z_{b_{N-3}}) \right] \stackrel{z_{b_2}\to 0}{=} h_{b_2}(h_{b_2}-1)f(z_{b_3},\dots,z_{b_{N-3}})
\end{equation}
and keeps the dependence on the $\bar{s}$-th copy of AdS space ($\bar{s}$-th particle) only via the leftover charge $Q_3^{[\bar{s}]}$ of the initial free wave at the AdS$_2$ boundary.

With the approach of this subsection, we can interpret all of the comb-channel conformal blocks and many cases with other topologies as limits of free-particle wavefunctions in tensor products of AdS. 
We will complete the discussion for any topology in the next subsection.

\subsubsection{Identity limits}
\begin{figure}[htp]
    \centering
    \includegraphics[scale=0.72]{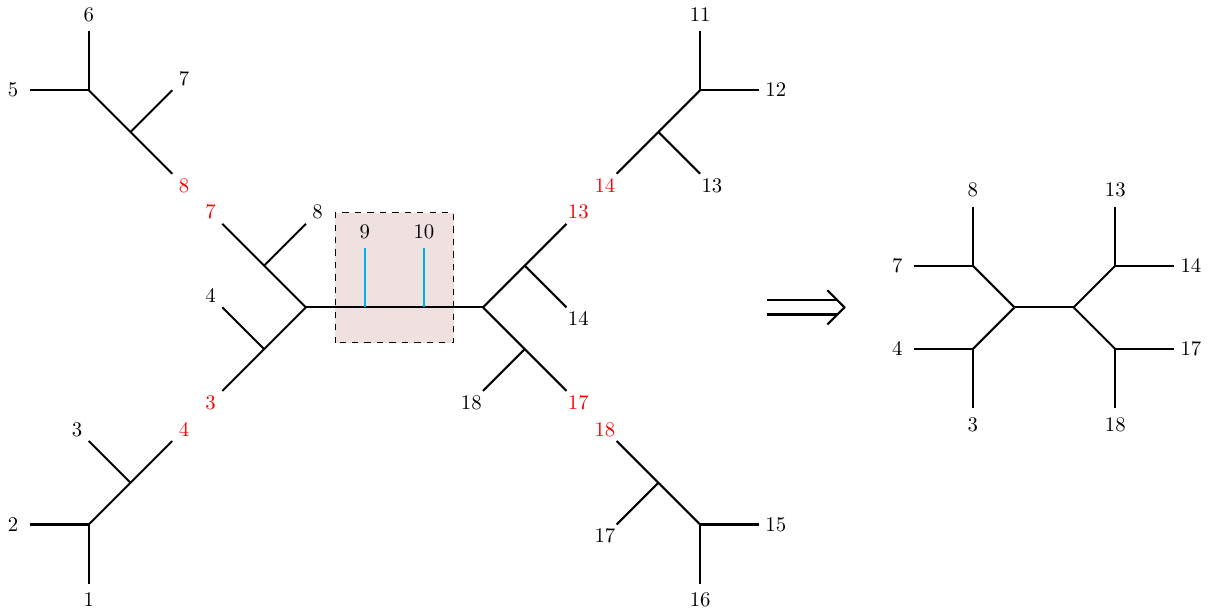}
    \caption{Example of how the most symmetric eight-point diagram can be seen as a limit of an 18-point 6PC diagram. Importantly, while one can perform either OPE or identity limits for the external combs, it is necessary to take identity limits for the legs in the highlighted part of the diagram.}
    \label{fig:eight-point-reduction}
\end{figure}

We saw in Section~\ref{ssec:identity_reduction} a very general way to reduce $N$-point conformal blocks to lower-point ones: the replacement of an external operator $\phi_k(x_k)$ with an identity operator $\mathds{1}(x_k)$, named in short \emph{identity limit}. These are crucial in our analysis since not all of the conformal blocks have topologies that can be interpreted as OPE limits of 6PC diagrams. For example, the eight-point function in the right part of Figure~\ref{fig:eight-point-reduction} has a topology that cannot be recovered by shortening combs that sit at the edges of the diagram: the two legs in the highlighted part of the diagram, in fact, are not directly coupled via an OPE to another external field, and can only be eliminated with a different procedure. 
For this purpose, the identity limits we introduced in Section~\ref{ssec:identity_reduction} are the perfect candidates.

Since whenever two external legs are paired in a three-point vertex we can use OPE limits, we will discuss for simplicity only identity limits for external legs that are attached to vertices with two internal legs.
Whenever we are taking such a limit, we are setting an external parameter $h_k$ that only appears in one of the conserved charges $Q_i^{[s]}$ to zero, and we are identifying the two Casimir operators that are attached to the leg $k$ where the identity is inserted. Since we start from a 6PC diagram---where external legs paired to two internal legs necessarily sit next to an AdS$_3$ Hamiltonian---this operation will always imply the tuning of a conserved charge to match the eigenvalue of the neighboring AdS$_3$ Hamiltonian:
\begin{equation}
    Q_i^{[s]}=\mathcal{H}^{[s]}\,.
\end{equation}
From the CFT picture, as we saw in Section~\ref{ssec:identity_reduction}, this reduction constrains the blocks to a codimension 1 surface described by $(N-4)$ cross-ratios $\zeta_r$ up to the ratio of prefactors $\tilde{\omega}_{\chan_N}=\Omega_{\chan_N}/\Omega_{\chan_{N-1}}$. Once this overall factor is stripped out, the leftover function corresponds to the $(N-1)$-point block $\tilde{\psi}^{(\chan_{N-1})}$.

Given our conventions introduced in Section~\ref{subsect:Conventions_Flow}, it's straightforward to work out the ratio of prefactors $\tilde{\omega}_{\chan_N}$ and the cross-ratios the lower-point functions will depend on. This information is enough to perform the reduction at the level of differential operators similar to our discussion of Section~\ref{subsec:example-identity-lim-6pt}. We will not provide a general rule on how to perform these identity limits, as these can involve more and more complicated resummations if there are long chains of vertices with three internal legs, but we display an explicit example in Appendix~\ref{app:11-point-limit} for the reduction of a twelve-point function to an eleven-point function via both OPE and identity limit. For general computational purposes, instead, we discuss in the next subsection some Feynman-like rules for one-dimensional conformal blocks.

\subsection{Feynman rules for one-dimensional conformal blocks}\label{SecRules}
Given the local nature of the OPE, which involves only operators that are paired in a three-point vertex within the OPE diagrams, and the presence of only scalar operators in one-dimensional CFTs---which implies additionally the absence of non-trivial tensor structures---it seems reasonable to expect conformal blocks to be constructible with some Feynman-like rules that involve just the individual three-point vertices. It was indeed shown in~\cite{Fortin:2020zxw} that this is the case, and a set of Feynman rules for all one-dimensional blocks were determined, although within conventions for OPE limits that differ from those of this paper, spelled out in Section~\ref{subsect:Conventions_Flow}.

Given the results of~\cite{Fortin:2020zxw}, then, it is reasonable to assume that an analogous set of Feynman rules has to hold within the conventions of this paper. This is corroborated by the existence of the gluing rules we described in Section~\ref{subsect:Gluing_six-pt} for 6PC diagrams, and the fact that any other OPE diagram can be obtained from limits of these. For further explicit checks, we refer the reader to Appendix~\ref{app:11-point-limit} and to the \texttt{Mathematica} notebook attached to this publication. We, therefore, aim in this section to spell out the Feynman rules that are valid in the conventions of this paper.

What we mean by the existence of Feynman rules, is that the one-dimensional conformal blocks can be written as
\begin{equation}
    \tilde{\psi}(z_1,\dots z_{N-3})=\sum_{n_1,\dots n_{N-3}}\prod_{i=1}^{N-3} \frac{z_i^{\mathbbm{h}_i+n_i}}{n_i!(2\mathbbm{h}_i)_{n_i}} \prod_{\nu=1}^{N-2} \chi_{\nu}
\end{equation}
where the $z_i$'s are determined according to the rules of Section~\ref{subsect:Conventions_Flow}, the index $\nu$ runs over the three-point vertices of the flow diagram (see Section~\ref{subsect:Conventions_Flow}), and for each vertex, a function $\chi_\nu$ is evaluated according to a fixed prescription that depends on the conformal data associated with the vertex $\nu$ of the flow diagram. 

The rules for the $\chi_\nu$ can be read off from the gluing coefficients $\alpha_{n_{b_i}}$, $\beta_{n_{b_i} n_{b_j}}$, and $\gamma_{n_{b_i}n_{b_j}n_{b_k}}$ we described in Section~\ref{subsect:Gluing_six-pt}, and depend on how many internal legs are involved in the vertex $\nu$:
\begin{enumerate}
    \item For vertices with one internal leg, we have the factor
\begin{equation}
    \chi_{h_{e_1},h_{e_2},\mathbbm{h}_{b}}=\left(\mathbbm{h}_{b}-h_{e_2}+h_{e_1} \right)_{n_b}\,,
    \label{eq:Feynman_one_internal}
\end{equation}
where $e_2$ is the conformal dimension associated with the field around which the OPE of this vertex is performed, \textit{i.e.}\ the label that gets passed to the neighboring vertex in the flow diagram.
\item For vertices with two internal legs the factor is
\begin{equation}
    \chi_{h_{e},\mathbbm{h}_{b_1},\mathbbm{h}_{b_2}}=\left(\mathbbm{h}_{b_1}+\mathbbm{h}_{b_2}-h_{e} \right)_{n_{b_1}+n_{b_2}}
\end{equation}
which can be seen to straightforwardly reduce to~\eqref{eq:Feynman_one_internal} in the $z_{b_1}\to 0$ or $z_{b_2}\to 0$ OPE limits.
\item Finally, for vertices with three internal legs we have the factor
\begin{multline}
    \chi_{\mathbbm{h}_{b_1},\mathbbm{h}_{b_2},\mathbbm{h}_{b_3}}=\left(\mathbbm{h}_{b_1b_2;b_3}-n_{b_3}\right)_{n_{b_1}}\left(\mathbbm{h}_{b_2b_3;b_1}-n_{b_1}\right)_{n_{b_2}}\left(\mathbbm{h}_{b_3b_1;b_2}-n_{b_2}\right)_{n_{b_3}}\\
    \times {}_4F_3\left(\begin{array}{c}
        -n_{b_1}, \, -n_{b_2},\, -n_{b_3},\, \mathbbm{h}_{b_1}+\mathbbm{h}_{b_2}+\mathbbm{h}_{b_3}-1  \\
        \mathbbm{h}_{b_1b_2;b_3}-n_{b_3}\,,\, \mathbbm{h}_{b_2b_3;b_1}-n_{b_1},\mathbbm{h}_{b_3b_1;b_2}-n_{b_2} 
    \end{array}; 1
    \right),
    \label{eq:FR_three_internal}
\end{multline}
where the internal legs are disposed in the clockwise order $(b_1,b_2,b_3)$ within the flow diagram.
\end{enumerate}

\noindent To furnish an efficient mean of computation for general conformal blocks, we attached to this publication a \texttt{Mathematica} notebook where, once an OPE diagram is provided as input in the form of a graph, there are functions that automatically compute Casimir equations and conformal block expressions from Feynman rules.

\section{Conclusions}\label{SecConc}

In this paper, we studied one- (and by their factorization also two-) dimensional higher-point conformal blocks from the point of view of the differential equations they satisfy. The presence of quadratic Casimir differential operators equips cross-ratio space with some natural choices of a metric, constructed by analyzing the terms that are second-order in derivatives. By conveniently choosing the set of cross-ratios and the prefactor, we find that any $N$-point one-dimensional conformal block can be seen to correspond to either free-particle wavefunctions in AdS$_3^{\otimes (\lceil N/3\rceil-1 )}$, or a limit thereof. 

A crucial step in this proof was to establish that for $N$-point OPE diagrams that are \emph{six-point constructible}---built of copies of six-point combs according to the definition and gluing rules of Section~\ref{sec:six-point-constructible}---the associated conformal blocks diagonalize $(N-3)/3$ copies of the AdS$_3$ free Hamiltonian. The simplest examples of this are the six-point comb-channel blocks that we analyzed in Section~\ref{SecHamComb} which correspond to wavefunctions propagating freely in just one copy of AdS$_3$. 

For all other types of diagrams, we argued in Section~\ref{subsect:non_6PC_diagrams} how these can be derived from the free-particle wavefunctions described above by taking either OPE limits or identity limits. These effectively restrict the Hamiltonians on subspaces of the initial tensor product space, an operation that introduces effective background potentials in the lower-dimensional setup. The simplest examples of this are the five-point blocks, corresponding to wavefunctions on AdS$_2$ in the presence of a background magnetic field, and the four-point blocks, which correspond to one-dimensional wavefunctions in the presence of a P\"oschl-Teller potential~\cite{Isachenkov:2016gim}. It is relevant to underline that our results directly state that all of the integrable systems that are associated with one- and two-dimensional conformal blocks can be seen as originating from a free theory.

While the presence of AdS spaces when discussing CFTs may not be surprising, let us stress that the results of this paper are, as they stand, completely disconnected from the AdS/CFT framework. The AdS$_3^{\otimes m}$ spaces that we describe here are in fact directly constructed in cross-ratio space, as opposed to the AdS$_2$ that is physical-space dual to the CFT$_1$ we discussed. Nonetheless, it would be interesting to understand whether the results of this paper can have any application to AdS physics. One such direction concerns the interpretation of the conserved charges that correspond to the Casimir operators that do not take part in the construction of the free Hamiltonians, which appear naturally in CFT but are a rather unfamiliar quantity to consider in AdS.
Furthermore, as the various factors in AdS$_3^{\otimes m}$ could also be interpreted as the configuration space for $m$ distinct particles in AdS$_3$, it could be of interest to understand if there can be any interesting consequence in expanding functions of $m$ points in the bulk of AdS$_3$ in terms of $(3m+3)$-point conformal blocks.

If we just focus on the CFT perspective, our work introduced new expressions and Feynman rules for conformal blocks that compared to~\cite{Fortin:2020zxw} are more symmetric around the vertices with three internal legs, cf.~\eqref{eq:FR_three_internal}. This implies, for example, that expressions for snowflake conformal blocks have a manifest cyclic symmetry in our conventions.

Finally, as the ultimate scope would be to better understand and compute higher-point conformal blocks in higher dimensions, let us underline that the results of this paper describe part of cross-ratio space for CFTs in any dimension. This means that the global topology of cross-ratio space for $N$-point blocks has to be such that it can accommodate the AdS$_3^{\otimes m}$ we discussed in this paper (or subspaces thereof) as slices of the whole space.
Our short explorations of five-point conformal blocks in any dimension show that their five-dimensional cross-ratio space in Euclidean signature does not correspond to a maximally symmetric space, but can be charted by four-dimensional slices parametrized by one variable $w\in [0,1]$ (as defined in~\cite{Buric:2020dyz,Buric:2021kgy}), such that the $w=0$ and $w=1$ slices are both 
AdS$_2^{\otimes 2}$ spaces, while the other slices acquire a different structure. 
Despite the absence of maximal symmetry (at least for five points), it would nonetheless be interesting to understand if the higher-dimensional integrable model associated with conformal blocks can also be seen as originating from a free theory in some higher-dimensional space. If this were to be true, it would surely shed light on how to optimize the computation of conformal blocks in dimensions higher than two.

\section*{Acknowledgements}
This work was supported by NSERC (JFF and LQ), the US Department of Energy under grant DE-SC00-17660 (WS), and in part by the Young Faculty Incentive Fellowship from IIT Delhi (SP).


\appendix
\section{OPE and identity limits: 11-point block example}
\label{app:11-point-limit}

In this appendix, we will show one instance of a non-6PC diagram that is obtainable through both the OPE limit and the identity limit.  The OPE limit will directly give rise to the answer expected from the Feynman rules while the identity limit will first need to be massaged to reach the same result.


\subsection{Necessary identities}

Before proceeding, we will need several well-known identities of hypergeometric functions, including Euler's identity
\begin{equation}\label{EqEuler}
    {}_2F_1\left(\begin{array}{c}a,\,b\\c\end{array};x\right)=(1-x)^{c-a-b}{}_2F_1\left(\begin{array}{c}c-a,\,c-b\\c\end{array};x\right)\,,
\end{equation}
as well as
\begin{equation}\label{EqpFq1}
\begin{gathered}
    {}_2F_1\left(\begin{array}{c}-n,\,b\\c\end{array};1\right)=\frac{(c-b)_n}{(c)_n}\,,\\
    {}_3F_2\left(\begin{array}{c}-n,\,b,\,d\\d,\,e\end{array};1\right)=\frac{(d-b)_n}{(d)_n}{}_3F_2\left(\begin{array}{c}-n,\,b,\,e-c\\b-d-n+1,\,e\end{array};1\right)\,.
\end{gathered}
\end{equation}
Here and for the rest of this subsection on identities, it is understood that $n$ is a non-negative integer while the remaining parameters are real.  As such, we also have for the Pochhammer symbol the relation $(1-a-n)_n=(-1)^n(a)_n$.

Another important identity is
\begin{equation}\label{EqpFq}
    {}_{p+1}F_{q+1}\!\left(\begin{array}{c}c+n,\,a_1,\,\cdots,\,a_p\\c,\,b_1,\,\cdots,\,b_q\end{array};1\right)=\sum_{j\geq0}\frac{(-1)^j(-n)_j}{(c)_jj!}\frac{\prod_{r=1}^p(a_r)_j}{\prod_{s=1}^q(b_s)_j}{}_pF_q\!\left(\begin{array}{c}a_1+j,\,\cdots,\,a_p+j\\b_1+j,\,\cdots,\,b_q+j\end{array};1\right),
\end{equation}
which is the $x=1$ case of
\begin{align*}
    {}_{p+1}F_{q+1}\!\left(\begin{array}{c}c\!+\!n,\,a_1,\,\cdots,\,a_p\\c,\,b_1,\,\cdots,\,b_q\end{array};x\right)&\!=\frac{x^{1-c}}{(c)_n}\frac{d^n}{dx^n}x^{c+n-1}{}_pF_q\left(\begin{array}{c}a_1,\,\cdots,\,a_p\\b_1,\,\cdots,\,b_q\end{array};x\right)\\
    &\!=\sum_{j\geq0}\frac{(-1)^j(-n)_j}{(c)_jj!}\frac{\prod_{r=1}^p(a_r)_j}{\prod_{s=1}^q(b_s)_j}x^j{}_pF_q\left(\begin{array}{c}a_1+j,\,\cdots,\,a_p+j\\b_1+j,\,\cdots,\,b_q+j\end{array};x\right).\nonumber
\end{align*}
We note that the first equality is obtained by deriving each term of the implicit sum separately while the second equality comes from the binomial expansion of the derivatives acting on the monomial and the hypergeometric function, as well as standard properties under derivation of the latter.


\subsection{Example: 11-point block}

\begin{figure}[htp]
    \centering
    \begin{minipage}{.4\textwidth}
    \includegraphics[scale=0.75]{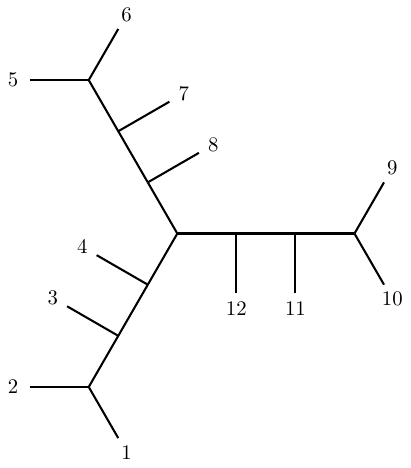}
    \end{minipage}
    \begin{minipage}{.05\textwidth}
    $\Rightarrow$
    \end{minipage}
    \begin{minipage}{.4\textwidth}
    \includegraphics[scale=0.75]{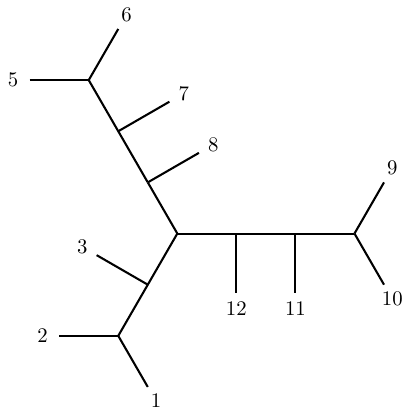}
    \end{minipage}
    \caption{6PC 12-point block (left) leading to the non-6PC 11-point block of interest (right) through either the identity limit, or an OPE limit combined with a redefinition of coordinates.}
    \label{fig:twelve-to-eleven}
\end{figure}

As described in the core of the paper, a non-6PC 11-point block can be reached through limits of a 6PC diagram, which for this example is shown in Figure~\ref{fig:twelve-to-eleven}.

We are interested in the 11-point block that corresponds to the OPE limit $x_1\to x_2$ (followed by the redefinitions $x_2\to x_1$, $x_3\to x_2$ and $x_4\to x_3$), or equivalently to the identity limit $h_4\to0$, of the 12-point block of Figure~\ref{fig:twelve-to-eleven}.  Before writing down the blocks, we introduce the convenient notation
\begin{equation}
    (a,b,\cdots)_n=(a)_n(b)_n\cdots\,,
\end{equation}
to simplify the equations here and throughout this appendix.  From the Feynman rules of Section~\ref{SecRules}, the 12-point block is
\begin{align}\label{Eq12ptblock}
    \tilde{\psi}_\text{$12$-pt}&=\sum_{n_1,\cdots,n_9\geq0}\frac{z_1^{\h_1+n_1}z_2^{\h_2}z_3^{\h_3+n_3}z_4^{\h_4+n_4}z_5^{\h_5}z_6^{\h_6+n_6}z_7^{\h_7+n_7}z_8^{\h_8}z_9^{\h_9+n_9}}{n_1!n_3!n_4!n_6!n_7!n_9!(2\h_1)_{n_1}(2\h_3)_{n_3}(2\h_4)_{n_4}(2\h_6)_{n_6}(2\h_7)_{n_7}(2\h_9)_{n_9}}\nonumber\\
    &\times(\h_1+h_1-h_2,\h_1+\h_2-h_3)_{n_1}(\h_{36;9}-n_9,\h_2+\h_3-h_4)_{n_3}\nonumber\\
    &\times(\h_4+h_5-h_6,\h_4+\h_5-h_7)_{n_4}(\h_{69;3}-n_3,\h_5+\h_6-h_8)_{n_6}\nonumber\\
    &\times(\h_7+h_9-h_{10},\h_7+\h_8-h_{11})_{n_7}(\h_{39;6}-n_6,\h_8+\h_9-h_{12})_{n_9}\nonumber\\
    &\times{}_2F_1\left(\begin{array}{c}\h_1+\h_2-h_3+n_1,\,\h_2+\h_3-h_4+n_3\\2\h_2\end{array};z_2\right)\nonumber\\
    &\times{}_2F_1\left(\begin{array}{c}\h_4+\h_5-h_7+n_4,\,\h_5+\h_6-h_8+n_6\\2\h_5\end{array};z_5\right)\nonumber\\
    &\times{}_2F_1\left(\begin{array}{c}\h_7+\h_8-h_{11}+n_7,\,\h_8+\h_9-h_{12}+n_9\\2\h_8\end{array};z_8\right)\nonumber\\
    &\times{}_4F_3\left(\begin{array}{c}-n_3,\,-n_6,\,-n_9,\,\h_3+\h_6+\h_9-1\\\h_{69;3}-n_3,\,\h_{39;6}-n_6,\,\h_{36;9}-n_9\end{array};1\right)\,,
\end{align}
while the 11-point block is
\begin{align}\label{Eq11ptblock}
    \tilde{\psi}_\text{$11$-pt}&=\sum_{m_2,\cdots,m_9\geq0}\frac{\zeta_2^{\h_1}\zeta_3^{\h_2+m_3}\zeta_4^{\h_4+m_4}\zeta_5^{\h_5}\zeta_6^{\h_6+m_6}\zeta_7^{\h_7+m_7}\zeta_8^{\h_8}\zeta_9^{\h_9+m_9}}{m_3!m_4!m_6!m_7!m_9!(2\h_2)_{m_3}(2\h_4)_{m_4}(2\h_6)_{m_6}(2\h_7)_{m_7}(2\h_9)_{m_9}}\nonumber\\
    &\times(\h_{26;9}-m_9,\h_1+\h_2-h_3)_{m_3}(\h_4+h_5-h_6,\h_4+\h_5-h_7)_{m_4}\nonumber\\
    &\times(\h_{69;2}-m_3,\h_5+\h_6-h_8)_{m_6}(\h_7+h_9-h_{10},\h_7+\h_8-h_{11})_{m_7}\nonumber\\
    &\times(\h_{29;6}-m_6,\h_8+\h_9-h_{12})_{m_9}\nonumber\\
    &\times{}_2F_1\left(\begin{array}{c}\h_1+h_1-h_2,\,\h_1+\h_2-h_3+m_3\\2\h_1\end{array};\zeta_2\right)\nonumber\\
    &\times{}_2F_1\left(\begin{array}{c}\h_4+\h_5-h_7+m_4,\,\h_5+\h_6-h_8+m_6\\2\h_5\end{array};\zeta_5\right)\nonumber\\
    &\times{}_2F_1\left(\begin{array}{c}\h_7+\h_8-h_{11}+m_7,\,\h_8+\h_9-h_{12}+m_9\\2\h_8\end{array};\zeta_8\right)\nonumber\\
    &\times{}_4F_3\left(\begin{array}{c}-m_3,\,-m_6,\,-m_9,\,\h_2+\h_6+\h_9-1\\\h_{69;2}-m_3,\,\h_{29;6}-m_6,\,\h_{26;9}-m_9\end{array};1\right)\,,
\end{align}
with their respective cross-ratios $z_{i=1,\cdots,9}$ and $\zeta_{i=2,\cdots,9}$.

Taking the OPE limit $x_1\to x_2$ of the 12-point block \eqref{Eq12ptblock} and redefining $x_2\to x_1$, $x_3\to x_2$ and $x_4\to x_3$ straightforwardly gives the 11-point block \eqref{Eq11ptblock} expected from the Feynman rules.  Taking the identity limit is however not as straightforward.

First, rewriting the 12-point cross-ratios in the identity limit $h_4\to0$ of \eqref{Eq12ptblock} in terms of the 11-point cross-ratios,
\begin{equation}
\begin{gathered}
    z_1=\zeta_2(1-z_2)\,,\qquad z_3=\frac{\zeta_3(1-z_2)}{z_2+\zeta_3(1-z_2)}\,,\qquad z_4=\zeta_4\,,\qquad z_5=\frac{(z_2+\zeta_3(1-z_2))\zeta_5}{z_2+\zeta_3(1-\zeta_6)(1-z_2)}\,,\\
    z_6=\frac{z_2\zeta_6}{z_2+\zeta_3(1-\zeta_6)(1-z_2)}\,,\qquad z_7=\zeta_7\,,\qquad z_8=\zeta_8\,,\qquad z_9=\frac{(z_2+\zeta_3(1-z_2))\zeta_9}{z_2}\,,
\end{gathered}
\end{equation}
and taking into account the prefactor
\begin{equation}
    \tilde{\omega}=\frac{1}{(1-z_2)^{h_3}}\left(\frac{1-z_3}{1-z_3(1-z_6)}\right)^{h_8}\,,
\end{equation}
it is now easy to see that the two limits would match if $S_\text{OPE}=S_\text{Id}$, where
\begin{align}\label{EqSOPE}
    S_\text{OPE}&=\sum_{m_2,m_3,m_5,m_6\geq0}\frac{\zeta_2^{m_2}\zeta_3^{m_3}\zeta_5^{m_5}\zeta_6^{m_6}}{m_2!m_3!m_5!m_6!}\frac{(\h_1+h_1-h_2,\h_1+\h_2+m_3-h_3)_{m_2}}{(2\h_1)_{m_2}(2\h_2)_{m_3}(2\h_5)_{m_5}(2\h_6)_{m_6}}\nonumber\\
    &\times(\h_{26;9}-n_9,\h_1+\h_2-h_3)_{m_3}(\h_4+\h_5+n_4-h_7,\h_5+\h_6+m_6-h_8)_{m_5}\nonumber\\
    &\times(\h_{69;2}-m_3,\h_5+\h_6-h_8)_{m_6}(\h_{29;6}-m_6)_{n_9}\nonumber\\
    &\times{}_4F_3\left(\begin{array}{c}-m_3,\,-m_6,\,-n_9,\,\h_2+\h_6+\h_9-1\\\h_{69;2}-m_3,\,\h_{29;6}-m_6,\,\h_{26;9}-n_9\end{array};1\right)
\end{align}
and
\begin{align}\label{EqSId}
    S_\text{Id}&=\sum_{n_1,n_3,n_6\geq0}(1-z_2)^{\h_1+\h_2+n_1+n_3-h_3}z_2^{-n_3}\frac{\zeta_2^{n_1}\zeta_3^{n_3}\zeta_6^{n_6}}{n_1!n_3!n_6!}\left(1+\frac{1-z_2}{z_2}\zeta_3\right)^{\h_{59;2}-n_3+n_9-h_8}\nonumber\\
    &\times\left(1+\frac{1-z_2}{z_2}\zeta_3(1-\zeta_6)\right)^{-\h_5-\h_6-n_6+h_8}\frac{(\h_1+h_1-h_2,\h_1+\h_2-h_3)_{n_1}}{(2\h_1)_{n_1}(2\h_6)_{n_6}}\nonumber\\
    &\times(\h_{26;9}-n_9)_{n_3}(\h_{69;2}-n_3,\h_5+\h_6-h_8)_{n_6}(\h_{29;6}-n_6)_{n_9}\nonumber\\
    &\times{}_4F_3\left(\begin{array}{c}-n_3,\,-n_6,\,-n_9,\,\h_2+\h_6+\h_9-1\\\h_{69;2}-n_3,\,\h_{29;6}-n_6,\,\h_{26;9}-n_9\end{array};1\right){}_2F_1\left(\begin{array}{c}2\h_2+n_3,\,\h_1+\h_2+n_1-h_3\\2\h_2\end{array};z_2\right)\nonumber\\
    &\times{}_2F_1\left(\begin{array}{c}\h_4+\h_5+n_4-h_7,\,\h_5+\h_6+n_6-h_8\\2\h_5\end{array};\frac{(1+\frac{1-z_2}{z_2}\zeta_3)\zeta_5}{1+\frac{1-z_2}{z_2}\zeta_3(1-\zeta_6)}\right)
\end{align}
are partial sums in the blocks obtained from the OPE limit $x_1\to x_2$ (or directly from the rules) \eqref{Eq11ptblock} and the identity limit $h_4\to0$ of \eqref{Eq12ptblock}, respectively.  Our strategy for the proof is simple: we will first rewrite $S_\text{Id}$ as sums of powers of the cross-ratios and then manipulate their coefficients to re-express them as the coefficients of $S_\text{OPE}$.  Clearly, for the identity to be satisfied, \eqref{EqSId} must be independent of $z_2$.

\subsubsection{Proof: First part}

To proceed, we use Euler's identity \eqref{EqEuler} on the ${}_2F_1(\cdots;z_2)$ of \eqref{EqSId}, followed by explicit expansions of the three hypergeometric functions, with summation indices $k_1$, $m_5$ and $k$ for ${}_2F_1(\cdots;z_2)$, ${}_2F_1(\cdots;\cdots\zeta_5)$ and ${}_4F_3(\cdots;1)$, respectively.  We then combine the different powers of functions of cross-ratios and binomial-expand $1+\frac{1-z_2}{z_2}\zeta_3$ and $1+\frac{1-z_2}{z_2}\zeta_3(1-\zeta_6)$ with summation indices $r_1$ and $r_2$, respectively.  Finally, we binomial-expand the remaining functions $1-z_2$ and $1-\zeta_6$ with summation indices $s_1$ and $s_2$, respectively.  We thus obtain
\begin{align}\label{EqSId1}
    S_\text{Id}&=\sum_{\substack{n_1,n_3,n_6,k_1,m_5\geq0\\k,r_1,r_2,s_1,s_2\geq0}}(-1)^{r_1+r_2}z_2^{k_1-r_1-r_2+s_1-n_3}\frac{\zeta_2^{n_1}\zeta_3^{n_3+r_1+r_2}\zeta_5^{m_5}\zeta_6^{n_6+s_2}}{k!r_1!r_2!s_1!s_2!n_1!k_1!n_3!m_5!n_6!}\nonumber\\
    &\times(-\h_{59;2}+n_3-m_5-n_9+h_8)_{r_1}(\h_5+\h_6+m_5+n_6-h_8)_{r_2}\nonumber\\
    &\times(-r_1-r_2)_{s_1}(-r_2)_{s_2}\frac{(\h_1+h_1-h_2,\h_1+\h_2-h_3)_{n_1}}{(2\h_1)_{n_1}(2\h_2)_{k_1}(2\h_5)_{m_5}(2\h_6)_{n_6}}\nonumber\\
    &\times(\h_{26;9}-n_9)_{n_3}(\h_{69;2}-n_3,\h_5+\h_6-h_8)_{n_6}(\h_{29;6}-n_6)_{n_9}\nonumber\\
    &\times\frac{(-n_3,-n_6,-n_9,\h_2+\h_6+\h_9-1)_k}{(\h_{69;2}-n_3,\h_{29;6}-n_6,\h_{26;9}-n_9)_k}(-n_3,\h_2-\h_1-n_1+h_3)_{k_1}\nonumber\\
    &\times(\h_4+\h_5+n_4-h_7,\h_5+\h_6+n_6-h_8)_{m_5}\,.
\end{align}
Redefining $r_2\to r-r_1+s_2$ allows us to resum over $r_1$, leading to a ${}_2F_1(\cdots;1)$ which can be rewritten in terms of Pochhammer symbols with the help of Euler's identity \eqref{EqEuler} (with the integer $n$ being $r$).  We then shift $r\to r-s_2$ and $k_1\to k_1-s_1$ and define $n_1=m_2$, $r=m_3-n_3$ and $n_6=m_6-s_2$ to get
\begin{align}\label{EqSId2}
    S_\text{Id}&=\sum_{\substack{m_2,m_3,m_5,m_6\geq0\\k,k_1,s_1,s_2,n_3\geq0}}(-1)^{m_3-n_3}z_2^{k_1-m_3}\frac{\zeta_2^{m_2}\zeta_3^{m_3}\zeta_5^{m_5}\zeta_6^{m_6}}{k!(k_1-s_1)!s_1!s_2!n_3!m_2!(m_3-n_3)!m_5!(m_6-s_2)!}\nonumber\\
    &\times(\h_{26;9}+n_3+m_6-n_9)_{m_3-n_3-s_2}(-m_3+n_3,\h_5+\h_6+m_5+m_6-s_2-h_8)_{s_2}\nonumber\\
    &\times(-m_3+n_3)_{s_1}\frac{(\h_1+h_1-h_2,\h_1+\h_2-h_3)_{m_2}}{(2\h_1)_{m_2}(2\h_2)_{k_1-s_1}(2\h_5)_{m_5}(2\h_6)_{m_6-s_2}}\nonumber\\
    &\times(\h_{26;9}-n_9)_{n_3}(\h_{69;2}-n_3,\h_5+\h_6-h_8)_{m_6-s_2}(\h_{29;6}-m_6+s_2)_{n_9}\nonumber\\
    &\times\frac{(-n_3,-m_6+s_2,-n_9,\h_2+\h_6+\h_9-1)_k}{(\h_{69;2}-n_3,\h_{29;6}-m_6+s_2,\h_{26;9}-n_9)_k}(-n_3,\h_2-\h_1-m_2+h_3)_{k_1-s_1}\nonumber\\
    &\times(\h_4+\h_5+n_4-h_7,\h_5+\h_6+m_6-s_2-h_8)_{m_5}\,.
\end{align}
Before proceeding with the second step, we first resum over $s_1$, which leads to
\begin{equation}
    {}_3F_2\left(\begin{array}{c}-k_1,\,1-2\h_2-k_1,\,-m_3+n_3\\1+\h_1-\h_2-k_1+m_2-h_3,\,1-k_1+n_3\end{array};1\right)\,,
\end{equation}
up to Pochhammer symbols, on which we can use \eqref{EqpFq1} and re-express as a sum over $s_1$.  Shifting $k_1\to k_1+s_1$ allows us to resum over $s_1$ through the binomial identity, effectively replacing the power of $z_2$ by $\frac{z_2}{1-z_2}$.  Defining $k_1=m_3-n$ gives
\begin{align}\label{EqSId3}
    S_\text{Id}&=\sum_{\substack{m_2,m_3,m_5,m_6\geq0\\n,k,s_2,n_3\geq0}}(-1)^{m_3-n_3-n}(1-1/z_2)^n\frac{\zeta_2^{m_2}\zeta_3^{m_3}\zeta_5^{m_5}\zeta_6^{m_6}}{k!s_2!m_2!(m_3-n_3)!m_5!(m_6-s_2)!}\nonumber\\
    &\times(\h_{26;9}+n_3+m_6-n_9)_{m_3-n_3-s_2}(-m_3+n_3,\h_5+\h_6+m_5+m_6-s_2-h_8)_{s_2}\nonumber\\
    &\times\frac{(\h_1+\h_2+m_2-h_3)_{m_3-n}}{(n_3+n-m_3)!(m_3-n)!}\frac{(\h_1+h_1-h_2,\h_1+\h_2-h_3)_{m_2}}{(2\h_1)_{m_2}(2\h_2)_{m_3-n}(2\h_5)_{m_5}(2\h_6)_{m_6-s_2}}\nonumber\\
    &\times(\h_{26;9}-n_9)_{n_3}(\h_{69;2}-n_3,\h_5+\h_6-h_8)_{m_6-s_2}(\h_{29;6}-m_6+s_2)_{n_9}\nonumber\\
    &\times\frac{(-n_3,-m_6+s_2,-n_9,\h_2+\h_6+\h_9-1)_k}{(\h_{69;2}-n_3,\h_{29;6}-m_6+s_2,\h_{26;9}-n_9)_k}\nonumber\\
    &\times(\h_4+\h_5+n_4-h_7,\h_5+\h_6+m_6-s_2-h_8)_{m_5}\,,
\end{align}
which completes the first step of our proof.

Comparing $S_\text{Id}$ \eqref{EqSId3} with $S_\text{OPE}$ \eqref{EqSOPE} it is easy to see that the identity limit reproduces the correct result if
\begin{equation}\label{EqFHyper}
    F={}_4F_3\left(\begin{array}{c}-m_3,\,-m_6,\,-n_9,\,\h_2+\h_6+\h_9-1\\\h_{69;2}-m_3,\,\h_{29;6}-m_6,\,\h_{26;9}-n_9\end{array};1\right)\,,
\end{equation}
where
\begin{align}\label{EqF}
    F&=\sum_{n,k,s_2,n_3\geq0}\frac{(-1)^{m_3-n_3-n}(1-1/z_2)^n}{k!s_2!(n_3+n-m_3)!(m_3-n_3)!}\frac{m_3!m_6!}{(m_3-n)!(m_6-s_2)!}\nonumber\\
    &\times(-m_3+n_3)_{s_2}\frac{(2\h_2)_{m_3}(2\h_6)_{m_6}}{(2\h_2)_{m_3-n}(2\h_6)_{m_6-s_2}}\frac{(\h_1+\h_2-h_3)_{m_2+m_3-n}}{(\h_1+\h_2-h_3)_{m_2+m_3}}\nonumber\\
    &\times\frac{(\h_{26;9}+n_3+m_6-n_9)_{m_3-n_3-s_2}}{(\h_{26;9}+n_3-n_9)_{m_3-n_3}}\frac{(\h_{69;2}-n_3)_{m_6-s_2}}{(\h_{69;2}-m_3)_{m_6}}\nonumber\\
    &\times\frac{(\h_{29;6}-m_6+s_2)_{n_9}}{(\h_{29;6}-m_6)_{n_9}}\frac{(-n_3,-m_6+s_2,-n_9,\h_2+\h_6+\h_9-1)_k}{(\h_{69;2}-n_3,\h_{29;6}-m_6+s_2,\h_{26;9}-n_9)_k}\,.
\end{align}

\subsubsection{Proof: Second part}

We now turn to proving that \eqref{EqF}, which is a quadruple sum, is independent of $z_2$ and is expressible in terms of a single sum as in \eqref{EqFHyper}.  We start by expressing the sum over $s_2$ as Pochhammer symbols times the hypergeometric function
\begin{equation}
    {}_4F_3\left(\begin{array}{c}\h_{29;6}-m_6+n_9,\,-m_6+k,\,-m_3+n_3,\,1-2\h_6-m_6\\\h_{29;6}-m_6+k,\,1-\h_{69;2}-m_6+n_3,\,1-\h_{26;9}-m_3-m_6+n_9\end{array};1\right)\,,
\end{equation}
that we rewrite as a sum over $j$ of
\begin{equation}
    {}_3F_2\left(\begin{array}{c}-m_6+k+j,\,-m_3+n_3+j,\,1-2\h_6-m_6+j\\1-\h_{69;2}-m_6+n_3+j,\,1-\h_{26;9}-m_3-m_6+n_9+j\end{array};1\right)\,,
\end{equation}
with the help of \eqref{EqpFq}.  We then apply the identity \eqref{EqpFq1} on the ${}_3F_2$ and rewrite the resulting ${}_3F_2$ as an explicit sum with summation index $s_2$.  Shifting $n_3\to n_3-n+m$ we obtain
\begin{align}\label{EqF1}
    F&=\sum_{n,j,k,s_2,n_3\geq0}\frac{(-1)^{j-n_3}(1-1/z_2)^n}{j!k!s_2!n_3!(n-n_3-j-s_2)!}\frac{m_3!}{(m_3-n)!}\nonumber\\
    &\times\frac{(2\h_2+m_3-n)_n(2\h_6+m_6-j)_j}{(\h_1+\h_2-h_3+m_2+m_3-n)_n}(\h_{69;2}-m_3+n_9)_{s_2}\nonumber\\
    &\times\frac{(\h_{26;9}+m_3+m_6-n_9+n_3-n)_{n-n_3-j-s_2}}{(\h_{26;9}+m_3-n_9+n_3-n)_{n-n_3}}\nonumber\\
    &\times\frac{(-m_3+n-n_3,\h_2+\h_6+\h_9-1)_k(-m_6)_{k+j+s_2}(-n_9)_{k+j}}{(\h_{69;2}-m_3)_{k+j+s_2}(\h_{29;6}-m_6)_{k+j}(\h_{26;9}-n_9)_k}\,.
\end{align}
On this result, we sum over $n_3$ and get, up to Pochhammer symbols,
\begin{equation}
    {}_3F_2\left(\begin{array}{c}m_3-n+1,\,-n+j+s_2,\,\h_{26;9}+m_3-n_9-n\\m_3-n+1-k,\,\h_{26;9}+m_3+m_6-n_9-n\end{array};1\right)\,,
\end{equation}
and we use \eqref{EqpFq} to express it as a sum over $i$ of
\begin{equation}
    {}_2F_1\left(\begin{array}{c}-n+j+s_2+i,\,\h_{26;9}+m_3-n-n_9+i\\\h_{26;9}+m_3+m_6-n_9-n+i\end{array};1\right)\,.
\end{equation}
Since the summation over $i$ forces $-n+j+s_2+i\leq0$, we then use \eqref{EqpFq1} and eliminate the ${}_2F_1$ for Pochhammer symbols, after which we can resum over $i$ resulting in
\begin{align}\label{EqF2}
    F&=\sum_{n,j,k,s_2\geq0}\frac{(-1)^{k+j}(1-1/z_2)^n}{j!k!s_2!(n-j-s_2)!}\frac{m_3!}{(m_3-n-k)!}\frac{(2\h_2+m_3-n)_n(2\h_6+m_6-j)_j}{(\h_1+\h_2-h_3+m_2+m_3-n)_n}\nonumber\\
    &\times(\h_{69;2}-m_3+n_9)_{s_2}\,{}_3F_2\left(\begin{array}{c}-k,\,\h_{26;9}+m_3-n_9-n,\,-n+j+s_2\\m_3-n+1-k,\,\-m_6-n+j+s_2+1\end{array};1\right)\nonumber\\
    &\times\frac{(m_6)_{n-j-s_2}}{(\h_{26;9}+m_3-n_9-n)_n}\frac{(\h_2+\h_6+\h_9-1)_k(-m_6)_{k+j+s_2}(-n_9)_{k+j}}{(\h_{69;2}-m_3)_{k+j+s_2}(\h_{29;6}-m_6)_{k+j}(\h_{26;9}-n_9)_k}\,.
\end{align}
We use once again \eqref{EqpFq1} on the ${}_3F_2$ in \eqref{EqF2} and rewrite the final ${}_3F_2$ as an explicit sum with summation index $i$.  We then shift $j\to j-s_2$ and resum over $s_2$ to produce the hypergeometric function
\begin{equation}
    {}_3F_2\left(\begin{array}{c}-j,\,\h_{69;2}-m_3+n_9,\,-\h_{29;6}+m_6-k-j+1\\2\h_6+m_6-j,\,n_9-k-j+1\end{array};1\right)\,,
\end{equation}
which can be transformed with the help of \eqref{EqpFq1} into another ${}_3F_2$ that can be rewritten explicitly as a sum over $s_2$, giving
\begin{align}\label{EqF3}
    F&=\sum_{n,i,j,k,s_2\geq0}\frac{(-1)^n(1-1/z_2)^n}{i!j!k!s_2!(n-j)!}\frac{m_3!}{(m_3-n)!}\frac{(2\h_2+m_3-n)_n}{(\h_1+\h_2-h_3+m_2+m_3-n)_n}\nonumber\\
    &\times\frac{(-j)_{s_2}}{(\h_{26;9}+m_3-n_9-n)_n}\frac{(-k,\h_{26;9}+m_3-n_9-n)_i}{(\h_{26;9}-n_9)_i(-m_6+i+1)_{j-n}}\nonumber\\
    &\times\frac{(\h_2+\h_6+\h_9-1,-n_9)_{k+j-s_2}(-m_6)_{k+j}}{(\h_{69;2}-m_3,\h_{29;6}-m_6)_{k+j-s_2}}\,.
\end{align}
Shifting $j\to j+s_2$ allows us to resum over $s_2$, generating
\begin{equation}
    {}_2F_1\left(\begin{array}{c}-n+j,\,-m_6+j+k\\-m_6+i+j-n+1\end{array};1\right)=\frac{(i-k-n+1)_{n-j}}{(-m_6+i+j-n+1)_{n-j}}\,,
\end{equation}
due to \eqref{EqpFq1} and thus eliminating one sum.  By shifting $k\to k-j$ the sum over $j$ can be performed, leading to
\begin{equation}
    {}_2F_1\left(\begin{array}{c}-n,\,-k+i\\-k+i-n+1\end{array};1\right)=\frac{(-n+1)_n}{(-k+i-n+1)_n}\,,
\end{equation}
eliminating another sum.  We notice now the presence of $(-n+1)_n=(-1)^n(0)_n=\delta_{0,n}$ which effectively kills a third sum, getting rid of the $z_2$ dependence at the same time, such that \eqref{EqF3} becomes
\begin{align}\label{EqF4}
    F&=\sum_{i,k\geq0}\frac{(-k,\h_{26;9}+m_3-n_9)_i}{(\h_{26;9}-n_9)_ii!}\frac{(\h_2+\h_6+\h_9-1,-m_6,-n_9)_k}{(\h_{69;2}-m_3,\h_{29;6}-m_6)_kk!}\,.
\end{align}
At this point it is straighforward to write the sum over $i$ as a ${}_2F_1$ and use \eqref{EqpFq1} (with $n$ being $k$) to eliminate a fourth sum, finally proving that \eqref{EqF4} is equal to the ${}_4F_3$ hypergeometric function \eqref{EqFHyper}.


\bibliographystyle{nb}
\bibliography{Biblio}

\end{document}